\documentclass[aps,preprint]{revtex4-2}
\usepackage{amsmath}
\usepackage[english]{babel}
\usepackage{amssymb}
\usepackage{graphicx}
\usepackage{epsfig}
\usepackage{bm}
\usepackage{dcolumn}
\usepackage{color}

\usepackage{bbm}
\usepackage{soul}
\usepackage{verbatim}
\usepackage{ulem}

\definecolor{violet}{rgb}{0.62, 0.0, 1.0}

\def\gcoup{g}

\def\gcoups{g_s}
\def\gcoupr{g_{v}}

\def\gmatrix{{\cal G}}

\def\jmatrix{{\cal J}}

\def\jnomatrix{{J}}

\def\rhofunmat{\mathbb{R}}

\def\rhofun{R}

\def\qindex{\scriptscriptstyle{Q}}

\def\trmin{{\rm tr}}
\def\mf{{\mbox{\tiny MF}}}
\DeclareMathOperator*{\sumint}{%
\mathchoice%
  {\ooalign{$\displaystyle\sum$\cr\hidewidth$\displaystyle\int$\hidewidth\cr}}
  {\ooalign{\raisebox{.14\height}{\scalebox{.7}{$\textstyle\sum$}}\cr\hidewidth$\textstyle\int$\hidewidth\cr}}
  {\ooalign{\raisebox{.2\height}{\scalebox{.6}{$\scriptstyle\sum$}}\cr$\scriptstyle\int$\cr}}
  {\ooalign{\raisebox{.2\height}{\scalebox{.6}{$\scriptstyle\sum$}}\cr$\scriptstyle\int$\cr}}
}

\DeclareMathOperator\arctanh{arctanh}
\DeclareMathOperator\arccoth{arccoth}
\DeclareMathOperator\arcsinh{arcsinh}

\begin{document}

\title{\sc\Large{Charged pseudoscalar and vector meson masses
under strong magnetic fields in an extended NJL model}}

\author{J.P.\ Carlomagno$^{a,b}$, D.\ G\'omez Dumm$^{a,b}$, M.F.\ Izzo Villafa\~ne$^{c}$,
S.\ Noguera$^{d}$ and N.N.\ Scoccola$^{b,c,d}$}

\affiliation{$^{a}$ IFLP, CONICET $-$ Departamento de F\'{\i}sica, Fac.\ de Cs.\ Exactas,
Universidad Nacional de La Plata, C.C. 67, (1900) La Plata, Argentina}
\affiliation{$^{b}$ CONICET, Rivadavia 1917, (1033) Buenos Aires, Argentina}
\affiliation{$^{c}$ Physics Department, Comisi\'{o}n Nacional de Energ\'{\i}a At\'{o}mica,
Av.\ Libertador 8250, (1429) Buenos Aires, Argentina}
\affiliation{$^{d}$ Departamento de F\'{\i}sica Te\'{o}rica and IFIC, Centro Mixto
Universidad de Valencia-CSIC, E-46100 Burjassot (Valencia), Spain}

\begin{abstract}
The mass spectrum of $\pi^+$ and $\rho^+$ mesons in the presence of a static
uniform magnetic field $\vec B$ is studied within a two-flavor NJL-like
model. We improve previous calculations taking into account the effect of
Schwinger phases carried by quark propagators, and using an expansion of
meson fields in terms of the solutions of the corresponding equations of
motion for nonzero $B$. It is shown that the meson polarization functions
are diagonal in this basis. Our numerical results for the $\rho^+$ meson
spectrum are found to disfavor the existence of a meson condensate induced
by the magnetic field. In the case of the $\pi^+$ meson, $\pi$ - $\rho$
mixing effects are analyzed for the meson lowest energy state. The
predictions of the model are compared with available lattice QCD results.
\end{abstract}


\maketitle

\renewcommand{\thefootnote}{\arabic{footnote}}
\setcounter{footnote}{0}

\section{Introduction}

It is well known that the presence of a background magnetic field of
magnitude $|B|\gtrsim 10^{19}$~G has a large impact on the physics of
strongly interacting particles, giving rise to significant effects on both
hadron properties and QCD phase transition
features~\cite{Kharzeev:2012ph,Andersen:2014xxa,Miransky:2015ava}. Such huge
magnetic fields can be achieved in matter at extreme conditions, e.g.\ at
the occurrence of the electroweak phase transition in the early
Universe~\cite{Vachaspati:1991nm,Grasso:2000wj} or in the deep interior of
compact stellar objects like
magnetars~\cite{Duncan:1992hi,Kouveliotou:1998ze}. Moreover, it has been
pointed out that values of $|eB|$ ranging from $m_\pi^2$ to $15\,m_\pi^2$
($|B|\sim 0.3$ to $5\times 10^{19}$~G) can be reached in noncentral
collisions of relativistic heavy ions at RHIC and LHC
experiments~\cite{Skokov:2009qp,Voronyuk:2011jd}. Though these large
background fields are short lived, they should be strong enough to affect
the hadronization process, offering the amazing possibility of recreating a
highly magnetized QCD medium in the lab. From the theoretical point of view,
the study of strong interactions in the presence of a large magnetic field
includes several interesting phenomena, such as the chiral magnetic
effect~\cite{Kharzeev:2007jp,Fukushima:2008xe,Kharzeev:2015znc}, which
entails the generation of an electric current induced by chirality
imbalance, and the so-called magnetic
catalysis~\cite{Klevansky:1989vi,Gusynin:1995nb} and inverse magnetic
catalysis~\cite{Bali:2011qj,Bali:2012zg}, which refer to the effect of the
magnetic field on the size of quark-antiquark condensates and on the
restoration of chiral symmetry.

Yet another possible effect has been discussed in the past few years. It has
been claimed that, for a sufficiently large external magnetic field, one
could find a phase transition of the QCD vacuum into an electromagnetic
superconducting state. This transition could be produced at zero
temperature, driven by the emergence of quark-antiquark vector condensates
that carry the quantum numbers of electrically charged $\rho$
mesons~\cite{Chernodub:2010qx,Chernodub:2011mc}. The existence or not of
such a superconducting (anisotropic and inhomogeneous) QCD vacuum state is
presently an interesting subject of investigation, and still remains as an
open question~\cite{Braguta:2011hq,Hidaka:2012mz,Li:2013aa,
Liu:2014uwa,Andreichikov:2016ayj,Bali:2017ian,Cao:2019res,Cao:2021rwx}.

It is clear that the study of the properties of magnetized light hadrons, in
particular $\pi$ and $\rho$ mesons, comes up as a crucial task towards the
understanding of the above mentioned problems. In fact, this subject has
been addressed in several works in the context of various effective schemes
for QCD. These include e.g.\ Nambu-Jona-Lasinio (NJL)-like
models~\cite{Chernodub:2011mc,Fayazbakhsh:2013cha,Fayazbakhsh:2012vr,Avancini:2015ady,Zhang:2016qrl,
Mao:2017wmq,GomezDumm:2017jij,Wang:2017vtn,Liu:2018zag,
Coppola:2018vkw,Mao:2018dqe,Avancini:2018svs,Avancini:2016fgq,Avancini:2018svs,
Coppola:2019uyr,Cao:2019res,Ghosh:2020qvg,Avancini:2021pmi,Cao:2021rwx},
quark-meson models~\cite{Kamikado:2013pya,Ayala:2018zat}, chiral
perturbation theory
(ChPT)~\cite{Andersen:2012zc,Agasian:2001ym,Colucci:2013zoa}, hidden local
symmetry~\cite{Kawaguchi:2015gpt}, path integral
Hamiltonians~\cite{Orlovsky:2013gha,Andreichikov:2016ayj} and QCD sum
rules~\cite{Dominguez:2018njv}. In addition, results for the charged $\pi$
and $\rho$ meson spectra in the presence of background magnetic fields have
been obtained from lattice QCD (LQCD)
calculations~\cite{Bali:2011qj,Luschevskaya:2015bea,Brandt:2015hnz,
Luschevskaya:2016epp,Bali:2017ian,Ding:2020hxw}.

In this article we concentrate on the analysis of charged $\pi^+$ and
$\rho^+$ mesons, which turn out to get mixed in the presence of an external
magnetic field $\vec B$. Our analysis is carried out in the framework of a
two-flavor NJL-like quark
model~\cite{Vogl:1991qt,Klevansky:1992qe,Hatsuda:1994pi}; within a similar
context, the properties of magnetized neutral pseudoscalar and vector mesons
have been studied in Ref.~\cite{Carlomagno:2022inu}. It is worth noticing
that in the present case the calculations involving quark loops for nonzero
$B$ require some care due to the presence of Schwinger
phases~\cite{Schwinger:1951nm}. While these phases cancel out for neutral
mesons, in general they do not vanish when charged mesons are considered;
this has been shown explicitly in the case of charged pions in
Refs.~\cite{Coppola:2018vkw,Coppola:2019uyr}. At the same time,
instead of dealing with free charged $\pi^+$ or $\rho^+$ meson fields, at
the zero order one should consider the wavefunctions obtained as solutions
of the charged meson equations of motion in the presence of a constant
external magnetic field $B$. In fact, at the one-loop level, Schwinger
phases induce a breakdown of translational invariance in quark propagators,
which is compensated by the wavefunctions of the external $\pi^+$ or
$\rho^+$ mesons. It is seen that the charged meson polarization functions
are not diagonal for the standard plane wave states, while they become
diagonalized in the basis associated to the solutions of the corresponding
equations of motion for nonzero $B$. In addition, it is important to care
about the regularization of ultraviolet divergences, since the presence of
the external magnetic field can lead to spurious results, such as unphysical
oscillations of physical observables~\cite{Allen:2015paa,Avancini:2019wed}.
Here we use the so-called magnetic field independent regularization (MFIR)
scheme,~\cite{Menezes:2008qt,Avancini:2015ady,Avancini:2016fgq,Coppola:2018vkw},
which has been shown to be free from these effects and to reduce the
dependence of the results on model parameters~\cite{Avancini:2019wed}.
Concerning the effective coupling constants of the model, we consider both
the case in which these parameters are fixed and the case in which they
depend on the external magnetic field. This last possibility, inspired by
the magnetic screening of the strong coupling constant occurring for large
$B$~\cite{Miransky:2002rp}, has been previously explored in effective
models~\cite{Ayala:2014iba,Farias:2014eca,Ferreira:2014kpa,Endrodi:2019whh,
Sheng:2021evj} in order to reproduce the inverse magnetic catalysis effect
obtained in finite-temperature LQCD calculations.

{}From our calculations it is found that the energy of the $\rho^+$
meson fundamental state ---which corresponds to a Landau level $k=-1$---
does not show a large reduction for values of $eB$ up to 1~GeV$^2$, both in
the cases of fixed and $B$-dependent couplings. Hence our approach, which
improves upon previous two-flavor NJL model calculations that use a
plane-wave approximation for charged meson wavefunctions, disfavors the
existence of a charged vector meson condensate induced by the magnetic
field. On the other hand, we find that for nonzero $B$ the lowest energy
state for the $\pi^+$ state ---Landau level $k=0$--- gets mixed with the
corresponding $\rho^+$ state, this mixing being quantitatively significant
for $eB$ above 0.5~GeV$^2$.

The paper is organized as follows. In Sec.~II we introduce the theoretical
formalism used to obtain the masses of charged meson eigenstates. In
particular, we obtain $\pi^+$ and $\rho^+$ polarization functions for the
lowest Landau levels $k=-1$ and $k=0$. In Sec.~III we present and discuss
our numerical results, while in Sec.~IV we provide a summary of our work,
together with our main conclusions. We also include Appendixes A, B and C to
provide some formulae related with the formalism, as well as some technical
details of our calculations.

\section{Theoretical formalism}

\subsection{Effective Lagrangian and mean field gap equation}
\label{sect2a}

Let us start by considering the Euclidean action for an extended NJL
two-flavor model in the presence of an electromagnetic field. We have
\begin{eqnarray}
S_E & = & \int d^4x\ \bigg\{ \bar \psi(x) \left(- i\, \rlap/\!D + m_c \right)
\psi(x)
\nonumber \\
& &
- \gcoups \Big[ \left( \bar\psi(x)  \psi(x) \right)^2 + \left( \bar\psi(x) \, i\gamma_{5} \vec \tau \psi(x)\right)^2 \Big]
- \,\gcoupr \left( \bar\psi(x) \,
\gamma_\mu\vec{\tau}\,\psi(x)\right)^2 \bigg\}
\ , \label{lagrangian}
\end{eqnarray}
where $\psi = (u\ d)^T$, and $m_c$ is the
current quark mass, which is assumed to be equal for $u$ and $d$
quarks. The interaction between the fermions and the
electromagnetic field ${\cal A}_\mu$ is driven by the covariant
derivative
\begin{equation}
D_\mu\ = \ \partial_{\mu}-i\,\hat Q \mathcal{A}_{\mu}\ ,
\label{covdev}
\end{equation}
where $\hat Q=\mbox{diag}(Q_u,Q_d)$, with $Q_u=2e/3$ and $Q_d =
-e/3$, $e$ being the proton electric charge. We consider the
particular case in which one has a homogenous stationary magnetic
field $\vec B$ orientated along the 3,  or $z$, axis. Then,
choosing the Landau gauge, we have $\mathcal{A}_\mu = B\, x_1\,
\delta_{\mu 2}$.

Since we are interested in studying meson properties, it is convenient to
bosonize the fermionic theory, introducing scalar, pseudoscalar and vector
fields $\sigma$, $\vec \pi(x)$ and $\vec \rho_\mu(x)$, and integrating out
the fermion fields. The bosonized
Euclidean action can be written as 
\begin{eqnarray}
S_{\mathrm{bos}} & = & -\ln\det\mathcal{D} +\frac{1}{4\gcoups}
\int d^{4}x\ \Big[\sigma(x)\sigma(x)+
\vec{\pi}(x)\cdot\vec{\pi}(x)\Big]
+ \, \frac{1}{4\gcoupr} \int d^{4}x\ \vec\rho_\mu(x)\cdot\vec\rho_\mu (x) \ ,
\label{sbos}
\end{eqnarray}
with
\begin{equation}
\mathcal{D}_{x,x'} \ = \ \delta^{(4)}(x-x')\,\big[-i\,\rlap/\!D +
m_0 + \sigma(x) + i\,\gamma_5\, \vec \tau \cdot \vec \pi(x) +
\gamma_\mu \, \vec \tau \cdot \vec \rho_\mu(x) \big] \ , \label{dxx}
\end{equation}
where a direct product to an identity matrix in color space is
understood.

We proceed by expanding the bosonized action in powers of the fluctuations
of the bosonic fields around the corresponding mean field (MF) values. We
assume that the field $\sigma(x)$ has a nontrivial translational invariant
MF value $\bar\sigma$, while the vacuum expectation values of other bosonic
fields are zero. Thus, the MF action per unit volume is given by
\begin{equation}
\frac{S^{\mbox{\tiny MF}}_{\mathrm{bos}}}{V^{(4)}} \ =
\ \frac{\bar \sigma^2}{4 \gcoups} - \frac{N_c}{V^{(4)}}
\sum_{f=u,d} \int d^4x \, d^4x' \ \trmin_D\, \ln
\left(\mathcal{S}^{\mbox{\tiny MF},\,f}_{x,x'}\right)^{-1} \ ,
\label{seff}
\end{equation}
where $\trmin_D$ stands for the trace in Dirac space, and
$\mathcal{S}^{\mf,\,f}_{x,x'} = \big( \mathcal{D}^{\mf,\,f}_{x,x'}
\big)^{-1}$ is the MF quark propagator in the presence of the
magnetic field. As is well known, the explicit form of the
propagators can be written in different
ways~\cite{Andersen:2014xxa,Miransky:2015ava}. For convenience we
take the form in which $\mathcal{S}^{\mf,\,f}_{x,x'}$ is given by
a product of a phase factor and a translational invariant
function, namely
\begin{equation}
\mathcal{S}^{\mf,\,f}_{x,x'} \ = \ e^{i\Phi_f(x,x')}\,\int \dfrac{d^4 p}{(2\pi)^4}\
e^{ip\, (x-x')}\, \tilde S_p^f \ , \label{sfx}
\end{equation}
where $\Phi_f(x,x')=q_f B (x_1+x_1')(x_2-x_2')/2$ is the so-called
Schwinger phase.
Now $\tilde S_p^f$ can be expressed in the Schwinger
form~\cite{Andersen:2014xxa,Miransky:2015ava}
\begin{eqnarray}
\tilde S_p^f & = & \int_0^\infty d\tau\
\exp\!\bigg[-\tau\Big(M^2+p_\parallel^2+p_\perp^2\,\dfrac{\tanh(\tau B_f)}{\tau B_f} - i \epsilon\Big) \bigg] \nonumber\\
& & \times \, \bigg\{\big(M-p_\parallel \cdot
\gamma_\parallel\big)\,\big[1+i s_f \, \gamma_1 \gamma_2 \,
\tanh(\tau B_f)\big] - \dfrac{p_\perp \cdot \gamma_\perp}{\cosh^2
(\tau B_f)}  \bigg\}\ , \label{sfp_schw}
\end{eqnarray}
where we have used the following definitions. The perpendicular
and parallel gamma matrices are collected in vectors $\gamma_\perp
= (\gamma_1,\gamma_2)$ and $\gamma_\parallel =
(\gamma_3,\gamma_4)$, and, similarly, we have defined $p_\perp =
(p_1,p_2)$ and $p_\parallel = (p_3,p_4)$. Note that we are working
in Euclidean space, where $\{ \gamma_\mu, \gamma_\nu \} = -2
\delta_{\mu\nu}$. Other definitions in Eq.~(\ref{sfp_schw}) are
$s_f = {\rm sign} (Q_f B)$ and $B_f=|Q_fB|$. The limit $\epsilon
\rightarrow 0$  is implicitly understood.

The integral in Eq.~\eqref{sfp_schw} is divergent and has to be properly
regularized. As stated in the Introduction, we use here the magnetic field
independent regularization (MFIR) scheme: for a given unregularized quantity
that depends explicitly on $B$, the corresponding (divergent) $B \to 0$
limit is subtracted and then it is added in a regularized form. Thus, the
quantities can be separated into a (finite) ``$B=0$'' part and a
``magnetic'' piece. Notice that, in general, the ``$B=0$" part still depends
implicitly on $B$ (e.g.\ through the values of the dressed quark masses
$M$), hence it should not be confused with the value of the studied quantity
at vanishing external field. To deal with the divergent ``$B=0$'' terms we
use here a 3D cutoff regularization scheme. In the case of the
quark-antiquark condensates $\phi_f \equiv \langle \bar \psi_f \,\psi_f
\rangle$, $f=u,d$, we obtain
\begin{equation}
\phi_f^{\rm reg} \ = \ \phi^{0,\,\rm reg} \, + \, \phi_f^{\rm
mag} \ ,
\end{equation}
where
\begin{equation}
\phi^{0,\rm reg} = -N_c\, M \, I_{1}\ , \qquad\qquad
\phi_f^{\rm mag} = -N_c\, M \,I_{1f}^{\rm mag}\ . \label{phif}
\end{equation}
The expression of $I_1$ for the 3D cutoff regularization is given by
Eq.~\eqref{I1freg} of App.\ A, while the $B$-dependent function $I_{1f}^{\rm
mag}$ reads~\cite{Allen:2015paa,Klevansky:1989vi}
\begin{equation}
I_{1f}^{\rm mag} = \dfrac{B_f}{2\pi^2} \left[ \ln \Gamma(x_f) -
\left(x_f - \dfrac{1}{2}\right) \ln x_f + x_f -
\dfrac{\ln{2\pi}}{2} \right] \, , \label{i1}
\end{equation}
where $x_f=M^2/(2B_f)$. The corresponding gap equation, obtained from
$\partial S^{\mbox{\tiny MF}}_\mathrm{bos} /
\partial \bar{\sigma} =0$, can be written as
\begin{eqnarray}
 M & = & m_c - 2 \gcoups \left[ \phi_u^{\rm reg} +  \phi_d^{\rm reg}\right].
\label{gapeqs}
\end{eqnarray}

\subsection{Charged meson sector}
\label{neutral}

As expected from charge conservation, it is easy to see that the
contributions to the bosonic action that are quadratic in the fluctuations
of charged and neutral mesons decouple from each other. We consider here the
charged meson sector in the presence of the external magnetic field. A
detailed analysis of the neutral sector can be found in
Ref.~\cite{Carlomagno:2022inu}. For definiteness we explicitly analyze the case of
positively charged mesons, since the results for meson masses will not
depend on the charge sign. The corresponding contribution to the quadratic
action can be written as
\begin{equation}
S^{{\rm quad},+}_{\mathrm{bos}} \ = \ \frac{1}{2}
\sum_{M,M'}\int d^4x \, d^4 x' \,
 \delta M(x)^\dagger \, {\gmatrix}_{MM'}(x,x') \, \delta M'(x')\ ,
\end{equation}
where $M,M'$ are either $\pi^+$ or $\rho^+_\mu$, and
\begin{eqnarray}
{\gmatrix}_{MM'}(x,x') \ &=& \ \frac{1}{2 \gcoup_M}\,
\delta_{MM'} \, \delta^{(4)}(x-x') +\, {\jmatrix}_{MM'}(x,x')
\end{eqnarray}
where $\delta_{MM'}$ is an obvious generalization of the Kronecker
$\delta$, and the constants $g_M$ are given by $g_{\pi^+} = \gcoups$
and $g_{\rho_\mu^+} = \gcoupr$.
The polarization functions ${\jmatrix}_{MM'}(x,x')$ read
\begin{equation}
{\jmatrix}_{M M'} (x,x') \ = \ 2 N_c\, \trmin_D \bigg[
\mathcal{S}^{\mf,u}_{x,x'} \, \Gamma_{M'} \,
\mathcal{S}^{\mf,d}_{x',x} \, \Gamma_M \bigg]\ ,
\label{jotas}
\end{equation}
with $\Gamma_{\pi^+} =  i \gamma_5$ and $\Gamma_{\rho_\mu^+} = \gamma_\mu$.

Contrary to the neutral meson case, here the Schwinger phases do not cancel,
due to their different quark flavors. As a consequence, the polarization
functions in Eq.~(\ref{jotas}) are not translational invariant, and they do
not become diagonal when transformed to the momentum basis. Instead of using
the standard plane wave decomposition, to diagonalize the polarization
functions it is necessary to expand the meson fields in terms of a set of
functions $\mathbb{F}(x,\bar q)$, associated to the solutions of the
corresponding equations of motion in the presence of a constant magnetic
field $B$. For the charged pion field we have
\begin{equation}
\delta \pi^+(x) \ = \ \dfrac{1}{2\pi}\sum_k \int
\frac{{dq_2\,dq_3\,dq_4}}{(2\pi)^3} \ \mathbb{F}(x,\bar q)
\, \delta \pi^+(\bar q)\ , \label{Ritus}
\end{equation}
where we have defined $\bar q = (k,q_2,q_3,q_4)$. Here the index
$k$ is an integer that labels the so-called Landau modes
associated to the presence of the magnetic field. The functions $
\mathbb{F}(x,\bar q)$ are given by
\begin{equation}
\mathbb{F}(x,\bar q) \ = \ N_k \, e^{i ( q_2 x_2 + q_3 x_3 + q_4 x_4)} \,
D_k(r)\ , \label{Fq}
\end{equation}
where $D_k(x)$ are the cylindrical parabolic functions with the convention
$D_k(x)=0$ for $k<0$ (this implies that for charged pions $k=0,1,2,\dots$).
We have used the definitions $N_k= (4\pi B_e)^{1/4}/\sqrt{k!}$ and $r =
\sqrt{2 B_e}\,x_1-s\sqrt{2/B_e}\,q_2$, where $B_e = |Q_{\pi^+} B| =
|eB|$, $s = \mathrm{sign}(Q_{\pi^+} B)=\mathrm{sign}(B)$, $Q_{\pi^+} = Q_u
- Q_d = e$. It is not difficult to show that the functions in
Eq.~(\ref{Fq}) are solutions of Klein-Gordon equation corresponding to a
pseudoscalar meson of mass $m_\pi$ in the presence of constant magnetic
field when the corresponding on-shell condition $(2 k+1) B_e+ q_3^2 + q_4^2
+ m_\pi^2=0$ is fulfilled.

In the case of the charged $\rho$ mesons we introduce a new set of functions
$\rhofunmat_{\mu\nu}(x,\bar q)$, expanding the vector fields as
\begin{equation}
\delta\rho^+_\mu(x) \ = \  \dfrac{1}{2\pi}\sum_k \int\frac{{dq_2\,dq_3\,dq_4}}{(2\pi)^3}\ \rhofunmat_{\mu\nu}(x,\bar q) \,
\delta\rho_{\nu}^+(\bar q) \ .
\label{Ritusvec}
\end{equation}
The new functions are given by
\begin{equation}
\rhofunmat_{\mu\nu}(x,\bar q) \ = \ \sum_{\ell = -1}^1 \rhofun_{\ell}(x,\bar q)\,
\Delta^{(\ell)}_{\mu\nu}\ ,
\label{Gqa}
\end{equation}
where
\begin{equation}
\rhofun_{\ell}(x,\bar q) \ = \ N_{k-s\ell} \ e^{i (q_2 x_2 + q_3 x_3 + q_4
x_4)} \, D_{k-s\ell}(r)\ .
\label{Gqaa}
\end{equation}
Note that in order to have non-vanishing functions $\rhofun_{\ell}(x,\bar
q)$, the condition $k-s \, \ell \geq 0$ has to be satisfied. Given the
possible values of $\ell$ ($0,\pm 1$) and $s$ ($\pm 1$), it follows that for
charged rho mesons one has $ k = -1,0,1,\dots$. There is some freedom in the
election of $\Delta^{(\ell)}$ matrices, which is compensated by the choice
of the meson polarization vectors. The explicit form of the matrices used
here is given in App.~B, together with the corresponding polarization
vectors $\epsilon_\nu(\bar q ,a)$, in terms of which one can write
the fields $\delta\rho_{\nu}^+(\bar q)$. In that appendix it is also shown
that the functions in Eq.~(\ref{Gqa}) are solutions of the Proca equation
corresponding to a vector meson of mass $m_\rho$ in the presence of constant
magnetic field when the on-shell condition $(2 k+1) B_e+ q_3^2 + q_4^2 +
m_\rho^2=0$ is fulfilled.

For convenience, in what follows we introduce the shorthand notation
\begin{equation}
\sumint_{\bar q}\ \equiv \
\dfrac{1}{2\pi}\sum_{k=k_{\rm min}}^\infty
\int\frac{{dq_2\,dq_3\,dq_4}}{(2\pi)^3}\ ,
\label{notation2}
\end{equation}
where it is understood that $k_{\rm min} = -1$ (0) for rho (pion) meson
fields. Hence, in the previously introduced basis we have
\begin{equation}
S^{{\rm quad},+}_{\mathrm{bos}} \ = \ \frac{1}{2} \sum_{MM'}
\sumint_{\bar q, \bar q'}\
  \delta M(\bar q)^\dagger  \, {\gmatrix}_{MM'} (\bar q,\bar q') \,
  \delta M'(\bar q')\ ,
\end{equation}
with
\begin{equation}
{\gmatrix}_{MM'}(x,x') \ = \ \frac{1}{2 \gcoup_M}\,
\delta_{MM'} \, \hat \delta_{\bar q \bar q'} +\,
{\jmatrix}_{MM'}(\bar q, \bar q')\ ,
\end{equation}
where we have defined
\begin{equation}
\hat \delta_{\bar q \bar q'} \ \equiv \ (2\pi)^4\,
\delta_{kk'}\, \delta(q_2-q_2')\, \delta(q_3-q_3')\, \delta(q_4-q_4')\ .
\end{equation}
The polarization functions ${\jmatrix}_{MM'}$ in this basis are given by
\begin{eqnarray}
{\jmatrix}_{\pi^+\pi^+} (\bar q,\bar q') & = &  2 N_c \int
d^4x\, d^4x' \ \trmin_D \Big( \mathcal{S}^{\mf,u}_{x,x'} \, i
\gamma_5 \, \mathcal{S}^{\mf,d}_{x',x} \, i \gamma_5 \Big) \;
\left[ \mathbb{F}(x,\bar q)\right]^\ast\, \mathbb{F}(x',\bar q') \ ,
\nonumber \\
{\jmatrix}_{\rho^+_\nu\rho^+_\mu}(\bar q,\bar q') & = & 2 N_c
\int d^4x\, d^4x' \ \trmin_D \Big( \mathcal{S}^{\mf,u}_{x,x'} \,
\gamma_\alpha \, \mathcal{S}^{\mf,d}_{x',x} \, \gamma_\beta \Big)
\;
\left[ \rhofunmat_{\beta\nu}(x,\bar q)\right]^\ast\,
\rhofunmat_{\alpha\mu}(x',\bar q')\ ,
\nonumber \\
{\jmatrix}_{\pi^+\rho^+_\mu}(\bar q,\bar q') & = & 2 N_c \int
d^4x\, d^4x' \ \trmin_D \Big( \mathcal{S}^{\mf,u}_{x,x'}\, \gamma_\alpha
\, \mathcal{S}^{\mf,d}_{x',x}\, i \gamma_5 \Big) \;
\left[ \mathbb{F}(x,\bar q)\right]^\ast\, \rhofunmat_{\alpha\mu}(x',\bar q') \
,
\nonumber \\
{\jmatrix}_{\rho^+_\mu\pi^+}(\bar q,\bar q') & = & 2 N_c \int
d^4x\, d^4x' \ \trmin_D \Big( \mathcal{S}^{\mf,u}_{x,x'}\, i\gamma_5
\, \mathcal{S}^{\mf,d}_{x',x} \, \gamma_\alpha \Big) \;
\left[\rhofunmat_{\alpha\mu}(x,\bar q)\right]^\ast\, \mathbb{F}(x',\bar q') \ .
\label{chargedpol}
\end{eqnarray}

In previous works~\cite{Coppola:2018vkw,Coppola:2019uyr} it has been shown
that ${\jmatrix}_{\pi^+\pi^+}(\bar q,\bar q')$ is diagonal in $\bar
q$-space. Following similar procedures, it is possible to show, after some
lengthy calculations, that this also holds for the other polarization
functions in Eqs.~(\ref{chargedpol}). Namely, one can write
\begin{eqnarray}
{\jmatrix}_{\pi^+\pi^+} (\bar q,\bar q')\ = \ \hat \delta_{\bar
q\bar q'}\,  {\jnomatrix}_{\pi^+\pi^+} (\bar q)\ , \qquad\qquad
{\jmatrix}_{\rho^+_\nu\rho^+_\mu}(\bar q,\bar q') \ = \ \hat
\delta_{\bar q\bar q'}\, {\jnomatrix}_{\rho^+_\nu\rho^+_\mu} (\bar
q)\ , & & \nonumber
\\
{\jmatrix}_{\pi^+\rho^+_\mu} (\bar q,\bar q') \ = \ \hat
\delta_{\bar q\bar q'}\, {\jnomatrix}_{\pi^+\rho^+_\mu} (\bar q)\ ,
\qquad\qquad {\jmatrix}_{\rho^+_\mu\pi^+}(\bar q,\bar q') \ = \ \hat
\delta_{\bar q\bar q'}\,  {\jnomatrix}_{\rho^+_\mu\pi^+} (\bar q)\ . & &
\label{diag}
\end{eqnarray}
Moreover, we also obtain ${\jnomatrix}_{\rho^+_{\mu}\pi^+} (\bar q) \propto
(0,0,q_4,-q_3)$ and, as expected, $[{\jnomatrix}_{\rho^+_{\mu}\pi^+} (\bar
q)]^\ast = {\jnomatrix}_{\pi^+\rho^+_{\mu}} (\bar q)$.

We are interested in studying the meson masses, i.e., the energies of
the lowest lying meson states. These correspond to the Landau modes $k=-1$
and $k=0$. From Eqs.~(\ref{diag}) it can be immediately seen that for the
mode $k=-1$ only the polarization function
${\jmatrix}_{\rho^+_\nu\rho^+_\mu}$ is nonzero; thus, this mode corresponds
to the lowest energy charged rho meson, which does not get mixed with the
pion sector. In turn, for the Landau mode $k=0$ one gets the lowest energy
charged pion, which gets coupled to the $k=0$ rho meson. In what follows we
analyze these two modes in detail.

\subsubsection{$k=-1$ charged $\rho$ meson}

For $k=-1$ only ${\jnomatrix}_{\rho^+_\nu\rho^+_\mu} (\bar q)$ is relevant.
Moreover, as discussed in App.~B, in this case there is only one possible
polarization vector available for the $\rho^+$ field. For $B>0$ (i.e.,
$s=1$) this vector is given by $\epsilon_\mu(\bar q_{(-1)},
1)=(1,0,0,0)$, while for $B<0$ ($s=-1$) one has $\epsilon_\mu(\bar
q_{(-1)}, 1)=(0,1,0,0)$, with the notation $\bar q_{(k)}=(k,q_2,q_3,q_4)$.
Thus, to get rid of Lorentz indices we can calculate the function
$\jnomatrix_{\rho^+\!\rho^+}(-1,\Pi^2)$, defined by
\begin{eqnarray}
\jnomatrix_{\rho^+\!\rho^+}(-1,\Pi^2) \ = \
\left[\epsilon_\nu(\bar q_{(-1)}, 1)\right]^\ast
\, \jnomatrix_{\rho^+_\nu\rho^+_\mu} (\bar q_{(-1)})
\, \epsilon_\mu(\bar q_{(-1)}, 1)
\end{eqnarray}
where $\Pi^2$ is the square of the canonical momentum (see Eq.~(\ref{pi2})).
For the $k=-1$ mode of a charged rho meson, one has $\Pi^2 = q_3^2+q_4^2 -
B_e$. The function $J_{\rho^+\!\rho^+}(-1,\Pi^2)$ is ultraviolet
divergent, and has to be regularized. As in the previous section, we
consider the MFIR scheme, in which we subtract the corresponding expression
in the $B\to 0$ limit and then we add it in a regularized form. In this way
we get the regularized expression
\begin{eqnarray}
\jnomatrix^{\rm reg}_{\rho^+\!\rho^+}(-1,\Pi^2) \ = \
J^{0,\rm reg}_{\rho}(\Pi^2)
+ {\jnomatrix}_{\rho^+\!\rho^+}^{\rm mag} (-1,\Pi^2)\ .
\label{regrhoplus}
\end{eqnarray}
The function $\jnomatrix^{0,\rm reg}_{\rho}(q^2)$, regularized through a 3D
cutoff, is given in Eq.~(\ref{b0reg}). Notice that it has an implicit
dependence on the magnetic field, through the value of the constituent quark
mass $M$. On the other hand, the ``magnetic'' piece is found to be given by
\begin{eqnarray}
{\jnomatrix}_{\rho^+\!\rho^+}^{\rm mag}(-1,\Pi^2) & = &
-\frac{N_c}{4\pi^2}\,\int_0^\infty dz \int_{-1}^1 dv
\ e^{-z [M^2 + (1-v^2)\Pi^2/4]} \nonumber \\
& & \times \, \bigg\{ \frac{(1+t_u)\,(1+t_d)}{\alpha_+}
\,\Big[ M^2 + \frac{1}{z} - \frac{1-v^2}{4}\,(\Pi^2 + B_e)\Big]
e^{-z (1-v^2)B_e/4} \nonumber \\
& & \hspace{0.5cm} - \, \frac{1}{z}
\,\Big[ M^2 + \frac{1}{z} - \frac{1-v^2}{4}\,\Pi^2\Big]
\bigg\}\ .
\label{rhorhom1}
\end{eqnarray}
where we have used the definitions $t_u = \tanh\left[ (1-v) z B_u/2
\right]$, $t_d= \tanh\left[ (1+v) z B_d/2 \right]$, $\alpha_+ = t_u/B_u +
t_d/B_d + B_e \, t_u t_d/(B_u B_d)$. Notice that, being
$J_{\rho^+\!\rho^+}(-1,\Pi^2)$ a function of $\Pi^2$, our result is
explicitly invariant under boosts in the direction of the magnetic field.

The mass of the $k=-1$ charged rho meson can be found
as a solution of the equation
\begin{eqnarray}
\frac{1}{2 \gcoupr}\, +\, \jnomatrix^{\rm
reg}_{\rho^+\!\rho^+}(-1,-m_{\rho^+}^2)\ = \ 0\ ,
\end{eqnarray}
while the associated energy will be given by $E_{\rho^+} =
\sqrt{m_{\rho^+}^2 - B_e}$.

{}By looking at the expression of the function $\jnomatrix^{0,\rm
reg}_{\rho}(q^2)$ (see Eqs.~(\ref{b0reg}) and (\ref{ima2})), one could
expect that the ${\jmatrix}_{\rho^+_\nu \rho^+_\mu}$ polarization function
gets an imaginary part when $m_{\rho^+} > 2 M$ or, equivalently, when
$E_{\rho^+}^2 > 4 M^2 - B_e$. In fact, in the absence of the external
magnetic field, the value $m_\rho=2M$ represents a threshold for the
appearance of an absorptive part in the $\rho$ meson propagator. This well
known feature of the NJL model is associated to the possible decay of the
meson into a quark-antiquark pair, and arises from the lack of confinement
in this effective approach. For nonzero $B$, however, the actual threshold
has to occur when the energy of $k=-1$ meson state satisfies $E_{\rho^+} >
2M$, i.e., for $m_{\rho^+} > m_{\rm th}^{(-1)}$, with $m_{\rm th}^{(-1)} =
\sqrt{4 M^2 + B_e}$. What happens in the interval $2 M < m_{\rho^+} <
m_{th}^{(-1)}$ is that the aforementioned imaginary part cancels out with
another imaginary contribution arising from the last term in the curly
brackets in Eq.~(\ref{rhorhom1}), after a proper analytic extension (notice
that this term makes the integral divergent for $\Pi^2<-4M^2$). Details of
this calculation are given in App.~C.

It is important to mention that, taking into account the $\rho^+$
polarization vector, it is possible to see that the spin of the $\rho^+$ in
the $k=-1$ state satisfies $S_z=s$. In this way, the vector meson spin is
shown to be aligned with the magnetic field, as expected for a positively
charged meson in its lowest Landau mode.

\subsubsection{$k=0$ sector}

Let us consider now the $k=0$ Landau mode. In this case there are two
transverse independent polarization vectors $\epsilon_\mu(\bar q_{(0)},
\ell)$, $\ell =1,2$, whose expressions are given in Eq.~(\ref{polk0}).
Taking into account the general form general
${\jnomatrix}_{\rho^+_{\mu}\pi^+} (\bar q) \propto (0,0,q_4,-q_3)$, it is
seen that in this case $\left[ \epsilon_\mu(\bar q_{(0)},1)\right]^\ast\,
{\jnomatrix}_{\rho^+_\mu \pi^+}(\bar q_{(0)}) = 0$. Thus, the charged pions
only mix with one of the two possible charged rho meson polarization states.
As expected, it can be shown that this state corresponds to the spin
projection $S_z=0$. We define now
\begin{eqnarray}
{\jnomatrix}_{\rho^+\pi^+} (0,\Pi^2) &=& \left[ \epsilon_\mu(\bar q_{(0)},2)\right]^\ast\,
{\jnomatrix}_{\rho^+_\mu \pi^+}(\bar q_{(0)})\ ,
\nonumber \\
{\jnomatrix}_{\rho^+\rho^+} (0,\Pi^2) &=& \left[ \epsilon_\nu(\bar q_{(0)},2)\right]^\ast \,
{\jnomatrix}_{\rho^+_\nu\rho^+_\mu} (\bar q_{(0)}) \, \epsilon_\mu(\bar q_{(0)},2)\ .
\end{eqnarray}
Note that for $k=0$ one has $\Pi^2 = q_3^2+q_4^2 + B_e$, both for $\pi^+$
and $\rho^+$ states. Evaluating the integrals in Eq.~(\ref{chargedpol}), the
explicit expression of the mixing piece ${\jnomatrix}_{\rho^+\pi^+}
(0,\Pi^2)$ is found to be given by
\begin{eqnarray}
{\jnomatrix}_{\rho^+\pi^+} (0,\Pi^2) \ = \ -\, i\,\frac{M N_c}{4 \pi^2}\, \sqrt{\Pi^2 - B_e}
\,\int_0^\infty dz \int_{-1}^1 dv \; \frac{t_u-t_d}{\alpha_+}\ e^{-z [M^2 +
(1-v^2)(\Pi^2-B_e)/4]}\ .
\label{rhopi}
\end{eqnarray}
It is not difficult to see that, as expected, ${\jnomatrix}_{\rho^+\pi^+}
(0,\Pi^2)$ vanishes at $B=0$. Moreover, the integrals are finite and, thus,
no regularization is needed. On the other hand, both
${\jnomatrix}_{\rho^+\rho^+} (0,\Pi^2)$ and ${\jnomatrix}_{\pi^+\pi^+}
(0,\Pi^2)$ turn out to be divergent. Therefore, as in the $k=-1$ case, we
use the MFIR scheme to get the corresponding regularized quantities, which
can be written as
\begin{eqnarray}
{\jnomatrix}^{\rm reg}_{\pi^+\pi^+} (0,\Pi^2) &=& J^{0,\rm reg}_{\pi}(\Pi^2) +  {\jnomatrix}_{\pi^+\pi^+}^{\rm mag}
(0,\Pi^2)\ ,
\nonumber \\
{\jnomatrix}^{\rm reg}_{\rho^+\rho^+} (0,\Pi^2) &=& J^{0,\rm reg}_{\rho}(\Pi^2) +  {\jnomatrix}_{\rho^+\!\rho^+}^{\rm mag}
(0,\Pi^2)\ .
\end{eqnarray}
The expressions for $J^{0,\rm reg}_{\pi}(\Pi^2)$ and $J^{0,\rm
reg}_{\rho}(\Pi^2)$ are given in App.~A. In the case of the charged pion,
the expression of the ``magnetic'' piece ${\jnomatrix}_{\pi^+\pi^+}^{\rm
mag}(0,\Pi^2)$ has been previously obtained in
Refs.~\cite{Coppola:2018vkw,Coppola:2019uyr}. For the reader's convenience,
we also quote it here. One has
\begin{eqnarray}
{\jnomatrix}_{\pi^+\pi^+}^{\rm mag} (0,\Pi^2) &=&- \dfrac{N_c}{4\pi^2}
\int_0^\infty\! dz \int_{-1}^1 dv \;
\nonumber \\
&& \hspace{-2.2cm}
\bigg\{\left[ \dfrac{1-t_u \,t_d}{\alpha_+} \bigg( M^2 +\dfrac{1}{z} -\frac{1-v^2}{4} (\Pi^2-B_e) \bigg)
+ \dfrac{(1-t_u^2)\,(1-t_d^2)}{\alpha_+^{2}}\right]  e^{-z [M^2 + (1-v^2)(\Pi^2-B_e)/4]}
\nonumber \\
&& \hspace{-1.7cm}
 - \dfrac{1}{z} \bigg[ M^2 +\dfrac{2}{z} -\frac{1-v^2}{4} \Pi^2 \bigg]  e^{-z [M^2 + (1-v^2)\Pi^2/4]} \bigg\}\ .
\label{Jmagpipi}
\end{eqnarray}
On the other hand, for the quadratic $\rho^+$ term we find
\begin{eqnarray}
{\jnomatrix}_{\rho^+\!\rho^+}^{\rm mag} (0,\Pi^2) &=& -\dfrac{N_c}{4\pi^2}
\int_0^\infty\! dz \int_{-1}^1 dv \;
\nonumber \\
&& \hspace{-1.5cm}
\bigg\{ \left[ \dfrac{1-t_u \,t_d}{\alpha_+} \bigg( M^2  -\frac{1-v^2}{4} (\Pi^2-B_e) \bigg)
 + \dfrac{(1-t_u^2)\,(1-t_d^2)}{\alpha_+^{2}}\right]  e^{-z[M^2 + (1-v^2)(\Pi^2-B_e)/4]}
\nonumber \\
&& \hspace{-1.cm}
- \dfrac{1}{z} \bigg[ M^2 +\dfrac{1}{z} -\frac{1-v^2}{4} \Pi^2 \bigg] e^{-z [M^2 + (1-v^2)\Pi^2/4]} \bigg\}\ .
\label{Jmagrhorho}
\end{eqnarray}
Once again, the polarization functions depend on $\Pi^2$; therefore, they
are invariant under boosts in the direction of the magnetic field.

{}From the above expressions, the pole masses of the physical mesons $\tilde
\pi^+$ and $\tilde \rho^+$ for the $k=0$ mode can be obtained as solutions
of the equation
\begin{eqnarray}
\mbox{det} \left(%
\begin{array}{cc}
 1/(2 \gcoup_s) + {\jnomatrix}^{{\rm reg}}_{\pi^+\pi^+}(0,-m^2) & {\jnomatrix}_{\rho^+\pi^+} (0,-m^2) \\
{\jnomatrix}_{\rho^+\pi^+} (0,-m^2)^\ast  &  1/(2 \gcoup_v ) +  {\jnomatrix}^{{\rm reg}}_{\rho^+\rho^+}(0,-m^2) \\
\end{array}%
\right) \ = \ 0\ ,
\label{deter}
\end{eqnarray}
while the associated meson energies are $E = \sqrt{m^2 + B_e}$. As in the
previous case, it is important to determine which is the threshold for the
appearance of absorptive parts. As in the case of ${\jnomatrix}^{\rm
reg}_{\rho^+\rho^+} (-1,-m^2)$, the ``$B=0$'' terms ${\jnomatrix}^{0,{\rm
reg}}_{\pi^+\pi^+} (0,-m^2)$ and ${\jnomatrix}^{0,{\rm reg}}_{\rho^+\rho^+}
(0,-m^2)$ get an imaginary part when $m > 2M$ [see Eqs.~(\ref{b0reg}) and
(\ref{ima2})]. Once again, these imaginary parts get cancelled by imaginary
contributions arising from the last terms in the integrands of
Eqs.~(\ref{Jmagpipi}-\ref{Jmagrhorho}), after analytic continuation. On the
other hand, by looking at the exponentials in Eqs.~(\ref{rhopi}),
(\ref{Jmagpipi}) and (\ref{Jmagrhorho}) one might naively expect to have a
threshold at $E = \sqrt{m^2+B_e} = 2M$, above which convergence would be
lost. From the physical point of view, however, this cannot be the case. To
see this, let us consider a noninteracting $u\bar d$ pair in the presence of
a magnetic field $\vec B= B \hat z$, with $B>0$. The lowest energy state
with spin projection $S_z=0$ will correspond to the configuration
$u(S_z=+\frac{1}{2})\,\bar d (S_z=-\frac{1}{2})$, i.e.~the $u$ quark lying
in its lowest Landau level, and the $d$ quark in its first excited Landau
level. Recalling that the energy of a spin 1/2 fermion in the presence of
the magnetic field quantizes as $E=\sqrt{m^2+2kQB}$, $k=0,1,\dots$, the
lowest possible energy for the noninteracting $\bar ud$ system will be given
by $E_u+E_d = M_u+\sqrt{M_d^2+2B_e/3}$ (note that the alternative spin
assignment, i.e. $u (S_z=-\frac{1}{2})\, \bar d (S_z=+\frac{1}{2})$
corresponds to a state with higher energy). In fact, what happens in our
case is that the factors $(t_u-t_d)$ and $(1-t_ut_d)$ in the integrals
contribute with an additional exponential behavior that pushes the actual
threshold up to $E> M+\sqrt{M^2+2 B_e/3}$, or, equivalently, $m >
m_{th}^{(0)}$, with $m_{th}^{(0)}= \sqrt{ \big(M+\sqrt{M^2+2 B_e/3}\,
\big)^2-B_e}$, in agreement with the physical expectation. It is worth
remembering that $M$ grows with $B$~\cite{Carlomagno:2022inu}, preventing
for an imaginary value of the threshold mass $m_{th}^{(0)}$.

Once the masses are determined, the composition of the physical meson states
$|\tilde\pi^+\rangle$ and $|\tilde\rho^+\rangle$ is given by the
corresponding eigenvectors that diagonalize the matrix in Eq.~(\ref{deter})
for $m=m_{\pi^+}$ and $m=m_{\rho^+}$. Thus, the mass eigenstates can be
written in terms of coefficients $c^M_{M'}$ as
\begin{eqnarray}
|\tilde\pi^+\rangle & = & c^{\tilde\pi^+}_{\pi^+} \, |\pi^+\rangle  + c^{\tilde\pi^+}_{\rho^+} \, |\rho^+\rangle \ ,
\nonumber \\
|\tilde\rho^+\rangle & = & c^{\tilde\rho^+}_{\pi^+} \, |\pi^+\rangle  + c^{\tilde\rho^+}_{\rho^+} \, |\rho^+\rangle \ .
\label{composition}
\end{eqnarray}

\section{Numerical results}

In what follows we quote the numerical results for the quantities discussed
in the previous section. We choose here the same set of model parameters as
in Ref.~\cite{Carlomagno:2022inu}, viz.\ $m_c = 5.833$~MeV, $\Lambda =
587.9$~MeV and $g_s\Lambda^2 = 2.44$. For vanishing external field, this
parametrization leads to an effective quark mass $M=400$~MeV and a
quark-antiquark condensate $\phi^{0,{\rm reg}} = (-241\ {\rm MeV})^3$; in
addition, one obtains the empirical values of the pion mass and decay
constant in vacuum, namely $m_\pi=138$~MeV and $f_\pi=92.4$~MeV. Regarding
the vector couplings, we take $g_{v}= 2.651/\Lambda^2$, which leads to
$m_\rho=770$~MeV at $B=0$. The behavior of quark masses and quark-antiquark
condensates as functions of $B$ can be found in
Ref.~\cite{Carlomagno:2022inu}.

As mentioned in the Introduction, while local NJL-like models lead to
magnetic catalysis at zero temperature, they fail to reproduce the so-called
inverse magnetic catalysis effect observed from lattice QCD calculations for
finite temperature systems. One simple way of dealing with this problem is
to allow the model coupling constants to depend on the magnetic field. With
this motivation, we also explore the possibility of considering magnetic
field dependent four-fermion couplings. For definiteness, in the case of the
$B$ dependence of the coupling $g_s$ we adopt here the form proposed in
Ref.~\cite{Avancini:2016fgq}, viz.
\begin{equation}
g_s(B) \ = \ g_s\, {\cal F}(B)\ ,
\label{gdeb}
\end{equation}
where
\begin{equation}
 {\cal F}(B) =  \kappa_1 + (1-\kappa_1) \, e^{-\kappa_2\,
(eB)^2}\ ,
\label{fdeb}
\end{equation}
with $\kappa_1=0.321$, $\kappa_2=1.31$~GeV$^{-2}$. With this assumption, it
is found that the effective quark masses are less affected by the presence
of the magnetic field than in the case of a constant $g_s$; in fact, they show
a non-monotonous behavior for increasing $B$, resembling the results found
in Refs.~\cite{Endrodi:2019whh,Avancini:2021pmi}. On the other hand, the
zero-temperature magnetic catalysis effect, characterized by the growth of
quark-antiquark condensates with the magnetic field, is similar for both a
constant coupling and for a $B$ dependent $g_s$ as in
Eqs.~(\ref{gdeb}-\ref{fdeb})~\cite{Avancini:2016fgq}. In the case of the
vector coupling constant $g_v$, for consistency we also allow for some
dependence on $B$. Due to the common gluonic origin of $g_s$ and $g_v$, we
assume that both couplings get affected in the same way by the magnetic
field, hence we take $g_{v}(B) = g_{v}\ {\cal F}(B)$.

\subsection{$k=-1$ charged $\rho$ meson}

\begin{figure}[h]
\centering{}\includegraphics[width=0.7\textwidth]{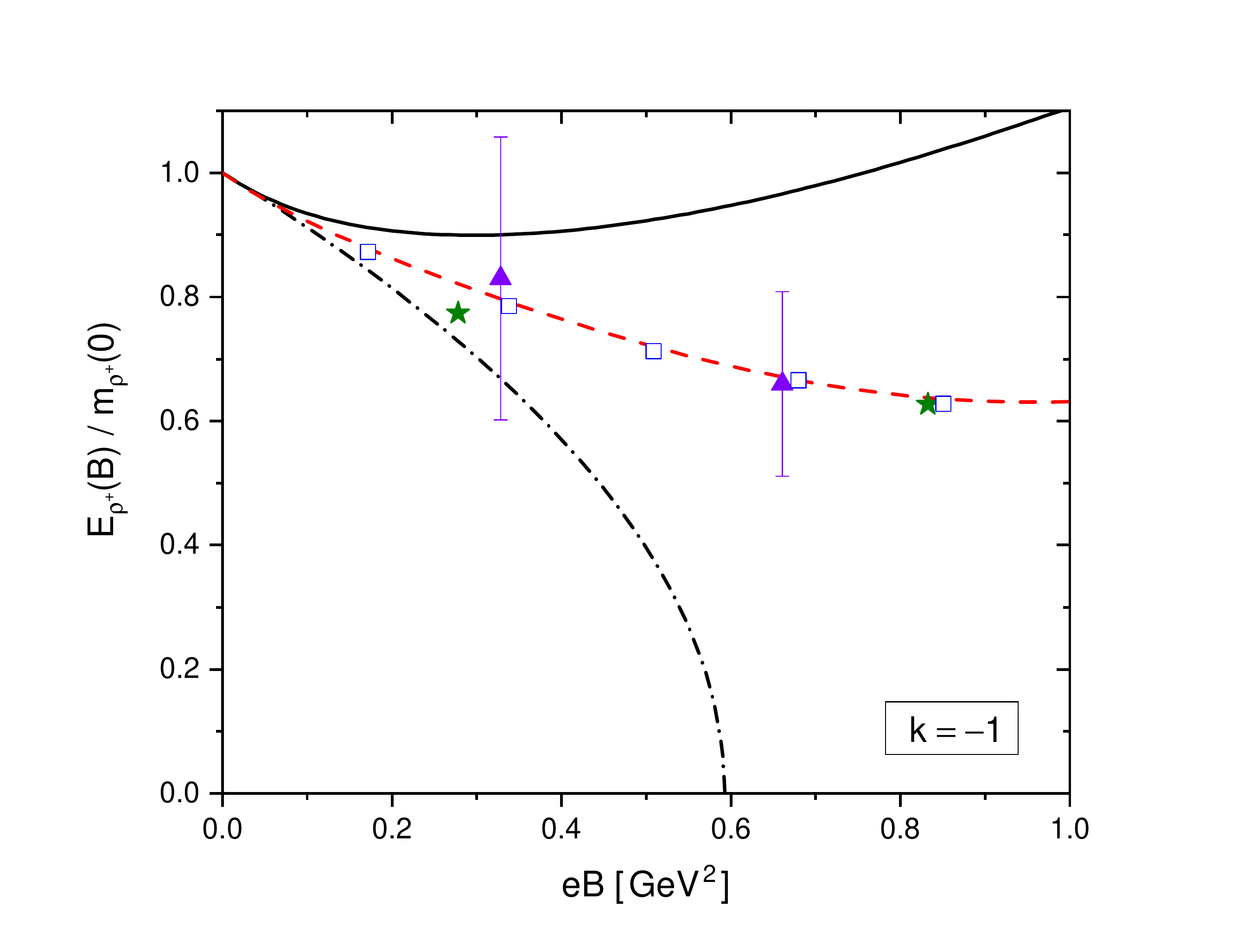}
\caption{(Color online) Energy of the $\rho^+$ meson as a function of $eB$
for the lowest Landau mode $k=-1$ and vanishing component of the momentum in
the direction of $\vec B$. Values are normalized to the $\rho^+$ mass at
zero external field. Solid and dashed lines correspond to fixed and
$B$-dependent coupling constants, respectively, while the dashed-dotted line
corresponds to a point-like $\rho^+$. For comparison, lattice QCD data
quoted in Refs.~\cite{Bali:2017ian}, \cite{Andreichikov:2016ayj} and
\cite{Hidaka:2012mz} are also included; they are indicated by triangles,
squares and stars, respectively.} \label{Fig1}
\end{figure}

In Fig.~\ref{Fig1} we show the energy of the $\rho^+$ meson, $E_{\rho^+}$,
as a function of the magnetic field, for the Landau mode $k=-1$ and
vanishing component of the $\rho^+$ momentum in the direction of $\vec B$.
The values are normalized to the energy at $B=0$, i.e.\ to the $\rho$ meson
rest mass $m_{\rho^+}(0) = E_{\rho^+}(0)$. As it has been extensively
discussed in the literature, if one takes the charged rho meson as a
point-like particle the energy behaves as $E_{\rho^+}(B) =
\sqrt{m_{\rho^+}^2 - eB}$, where $m_{\rho^+}$ is a constant mass. This leads
to a strong decrease with the magnetic field (dashed-dotted line in
Fig.~\ref{Fig1}) that reaches $E_{\rho^+}(B) = 0$ at $eB \sim 0.6$~GeV$^2$,
triggering the appearance of a charged vector meson condensate.

As shown in Fig.~\ref{Fig1}, it is found that our results do not support the
existence of this condensate. The full line in the figure corresponds to the
normalized energy for the case in which the four-fermion coupling constants
$g_s$ and $g_v$ are kept fixed. We see that although for low values of $eB$
the $\rho^+$ energy shows a decreasing behavior, at $eB \sim
0.2-0.3$~GeV$^2$ the curve reaches a minimum, and for larger values of the
magnetic field the energy gets steadily increased. In the case in which the
four-fermion couplings are taken to be dependent on $B$ (red dashed line),
the situation appears to be qualitatively similar, although the minimum is
found at a larger value $eB \sim 0.9$~GeV$^2$. Therefore, in both situations
the model does not predict the presence of $\rho^+$ condensation within the
considered range of values of $eB$. This behavior is in general consistent
with the results obtained through LQCD calculations; for comparison, in
Fig.~\ref{Fig1} we include LQCD data taken from Refs.~\cite{Bali:2017ian},
\cite{Andreichikov:2016ayj} and \cite{Hidaka:2012mz}, indicated by
triangles, squares and stars, respectively.

It is worth mentioning that our results differ substantially from those
obtained in other works in the framework of two-flavor NJL-like
models~\cite{Liu:2014uwa,Cao:2019res}, which do find $\rho^+$ meson
condensation for $eB\sim 0.2$ to 0.6~GeV$^2$. In those works it is assumed
that charged pions and vector mesons lie in zero three-momentum states. Here
we use, instead, an expansion of meson fields in terms of the solutions of
the corresponding equations of motion for nonzero $B$ [see
Eqs.~(\ref{Ritus}-\ref{Gqaa})], taking properly into account the presence of
Schwinger phases in quark propagators. Our numerical analysis shows that
this has a dramatic incidence in the numerical results, implying a
qualitative change in the behavior of the $\rho^+$ energy for the $k=-1$
Landau mode.

\subsection{$k=0$ sector}

In this subsection we present and discuss the results associated with the
$k=0$ sector. As in Sec.~II.B, we will concentrate on the subsystem that
contains the lowest energy pion state, i.e.\ the one formed by $\pi^+$ and
$\rho^+$ states with polarization $\epsilon_\nu(\bar q_{(0)} ,2)$ [see
Eq.~(\ref{polk0})], corresponding to a spin projection $S_z=0$. As stated,
the mass eigenstates denoted by $\tilde\pi^+$ and $\tilde\rho^+$ are
obtained as combinations of the states $\pi^+$ and $\rho^+$ in
Eq.~(\ref{sbos}). Here $\tilde\pi^+$ and $\tilde\rho^+$ are expected to be
the states with lower and higher energies, respectively.

\begin{figure}[h]
\centering{}\includegraphics[width=1\textwidth]{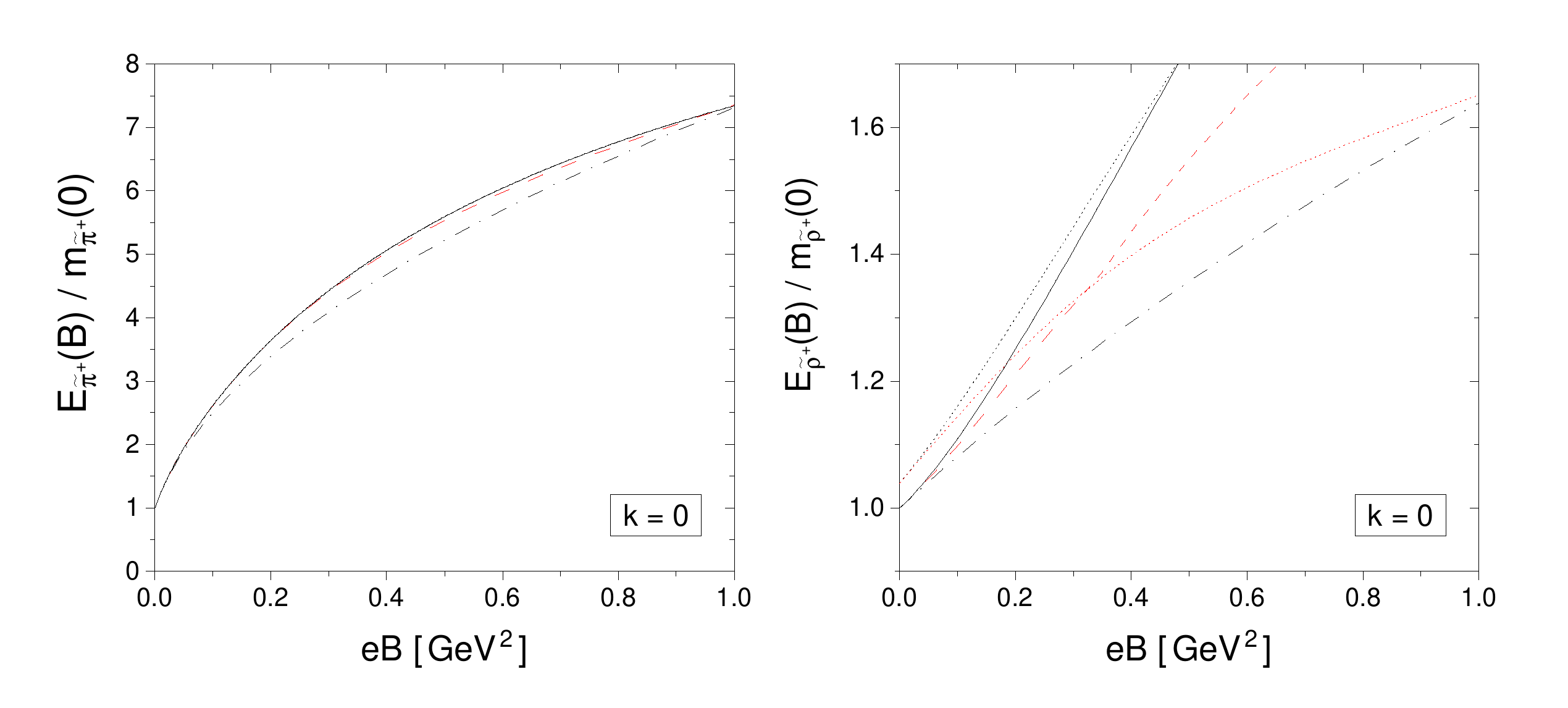}
\caption{(Color online) Energy of the $\tilde\pi^+$ (left) and
$\tilde\rho^+$ (right) mass eigenstates as functions of $eB$, for the Landau
mode $k=0$ and vanishing components of the momenta in the direction of $\vec
B$. Values are normalized to the meson masses at zero external field. Solid
and dashed lines correspond to fixed and $B$-dependent coupling constants,
respectively, while dashed-dotted lines correspond to the cases in which the
mesons are assumed to be point-like. In the right panel, the dotted lines
indicate the thresholds for the decay of the $\tilde\rho^+$ meson into a
$u\bar d$ pair. In the case of $B$-dependent couplings, the estimated energy
beyond this threshold is shown by the red short-dashed line.} \label{Fig2}
\end{figure}

The energies of the mass eigenstates as functions of the external magnetic
field, normalized to the values of the corresponding masses at $B=0$, are
shown in Fig.~\ref{Fig2}.
In the left panel we display the results for the $\tilde\pi^+$ state; the
full black line corresponds to the case in which the four-fermion couplings
are kept fixed, while the red dashed line indicates the relative
$\tilde\pi^+$ energy when $g_s$ and $g_v$ depend on $B$ in the form given by
Eq.~(\ref{fdeb}). As a reference, the behavior of $E_{\pi^+}(B)/m_{\pi^+}(0)$
for a point-like pion is also shown (black dash-dotted line). From the
figure it can be seen that our results for the $\tilde\pi^+$ state are
almost independent on whether the four-fermion couplings are taken to be
constant or not. Within the considered range of values of $eB$, in both
cases the energy shows a monotonous increasing behavior that goes slightly
above the one obtained for the point-like particle approximation. Our
results for the $\tilde\rho^+$ energy are given in the right panel of
Fig.~\ref{Fig2}, where the same line convention is used.  We see that in
this case the values are somewhat more sensitive on whether the four-fermion
couplings are taken to be constant or not. In both cases the energy shows an
increasing behavior, which is found to be steeper than the one obtained in
the point-like particle approximation. We also include in the graph the
corresponding thresholds for the decay of the $\tilde\rho^+$ into a $u\bar
d$ pair (thin short-dotted lines). If the couplings $g_s$ and $g_v$ are kept
constant, we see that the $\tilde\rho^+$ energy lies below the threshold for
the range plotted in the figure. On the other hand, in the case of
$B$-dependent couplings the corresponding threshold is reached at $eB \simeq
0.32$~GeV$^2$, $E_{\rho^+}\simeq 1.34\, m_{\rho^+}(0)$. For larger values of
the external field, the quark loop in the associated polarization function
includes an absorptive piece corresponding to an unphysical decay of the
$\tilde\rho^+$ meson into a quark-antiquark pair. As discussed in the
previous section, although in this region one can still obtain results for
the $\tilde\rho^+$ energy by means of an analytic extension of the
polarization function (short-dotted red curve in the right panel of
Fig.~\ref{Fig2}), these predictions have to be taken as merely indicative.

The composition of the mass eigenstates can be analyzed by looking at the
coefficients $c^M_{M'}$ introduced in Eq.~(\ref{composition}). The
corresponding results for some representative values of the magnetic field
are listed in Table~\ref{taba}. They correspond to the case in which the
four-fermion couplings are kept constant, and are similar to those obtained
in the case of $B$-dependent $g_s$ and $g_v$. We note that while the energies
do not depend on whether $B$ is positive or negative, the corresponding
eingenvectors do; the relative signs in Table~\ref{taba} correspond to the
choice $B>0$.
\begin{table}[h]
\begin{center}
\begin{tabular}{ccccccc}
\hline
\hline
 State         &$\ \ \ \ $ &  $eB\ [{\rm GeV}^2]$   &    & \hspace{3mm}
 $c^{\tilde\pi^+}_{\pi^+}$\hspace{3mm} & \hspace{3mm}$c^{\tilde\pi^+}_{\rho^+}$\hspace{3mm}
      \\
\hline
 $\tilde\pi^+$         & & 0.05 & &   0.999      &  0.013            \\
                           & & 0.5  & &   0.960           &  0.281               \\
                           & & 1.0  & &   0.892           &  0.453               \\
\hline
 State         &$\ $ &  $eB\ [{\rm GeV}^2]$   &    & \hspace{3mm}
 $c^{\tilde\rho^+}_{\pi^+}$\hspace{3mm} & \hspace{3mm}$c^{\tilde\rho^+}_{\rho^+}$\hspace{3mm}
      \\
\hline
 $\tilde\rho^+$        & & 0.05 & &   -0.156    &   0.988            \\
                                & & 0.5  & &  -0.702          &   0.713              \\
\hline
\end{tabular}
\caption{Composition of the $k=0$, $S_z = 0$ charged meson mass eigenstates
for some selected values of $eB$. Relative signs correspond to the choice
$B>0$.}
\label{taba}
\end{center}
\end{table}
As expected, for low magnetic fields (e.g.\ $eB=0.05$~GeV$^2$) the
eigenstates $\tilde \pi^+$ and $\tilde \rho^+$ are almost pure $\pi^+$ and
$\rho^+$, respectively, while the mixing gets increased as $eB$ grows. In
the case of the $\tilde \pi^+$ state, we find that the $\rho^+$ component
reaches a fraction of about $|c^{\tilde\pi^+}_{\rho^+}|^2=0.2$ (i.e., about a 20\%) at
$eB=1$~GeV$^2$. For the $\tilde \rho^+$ state the admixture grows faster
with $eB$, both $\pi^+$ and $\rho^+$ components having approximately equal
weight for $eB=0.5$~GeV$^2$ (i.e.\ close to the threshold for
quark-antiquark production, see the short-dotted black curve in the right
panel of Fig.~\ref{Fig2}).

\begin{figure}[h]
\centering{}\includegraphics[width=0.7\textwidth]{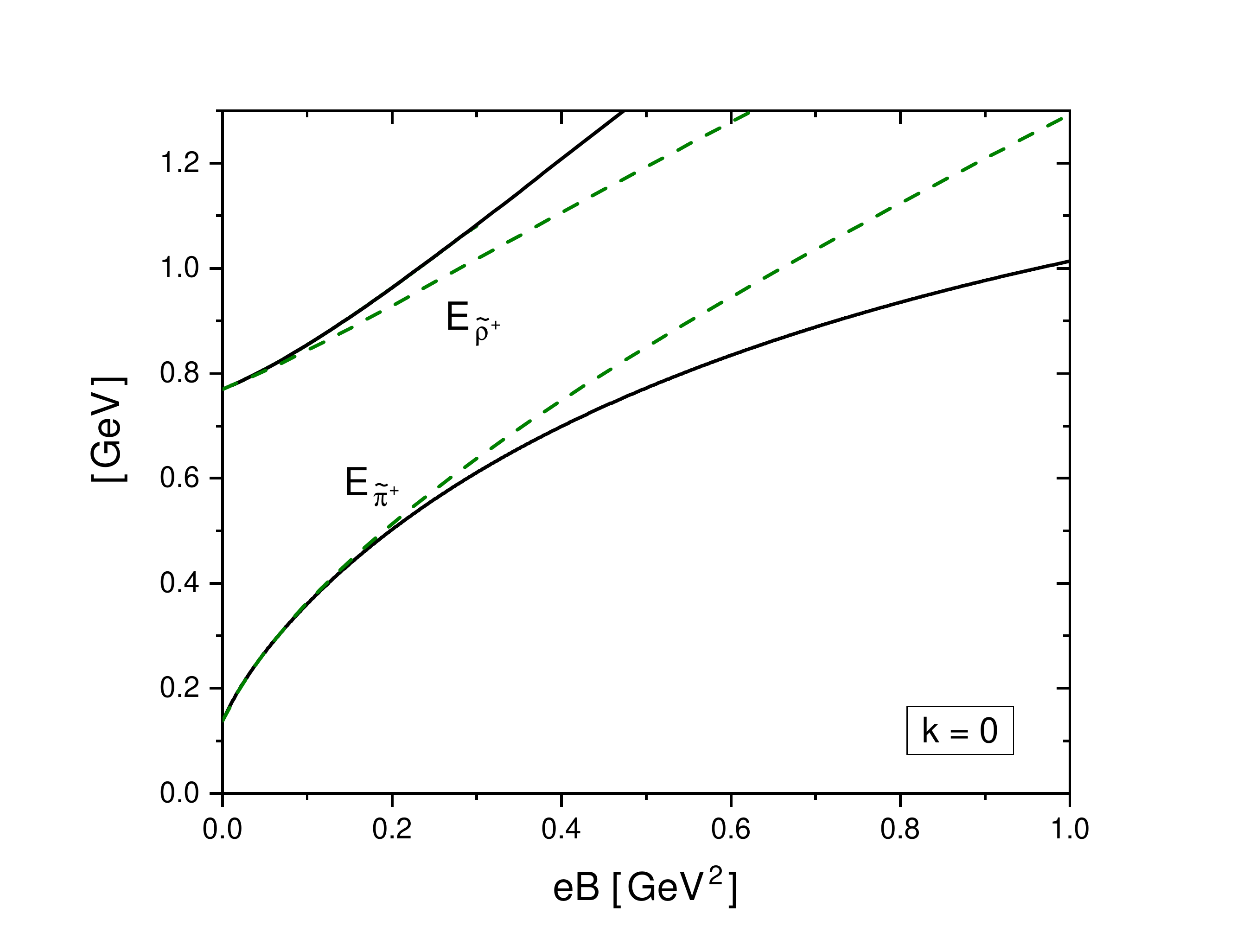}
\caption{(Color online) Energies of the $\tilde\pi^+$ and $\tilde\rho^+$
mass eigenstates as functions of $eB$, for the Landau mode $k=0$ and
vanishing components of the momenta in the direction of $\vec B$. Solid and
dashed lines correspond to the calculations with and without the inclusion
of the $\rho^+ - \pi^+$ mixing terms, respectively.} \label{Fig3}
\end{figure}

Let us now analyze the impact of the pseudoscalar-vector mixing on the
energies of the $\tilde \pi^+$ and $\tilde \rho^+$ states. In
Fig.~\ref{Fig3} we show the dependence of these energies on the magnetic
field, considering both the case in which the mixing is taken into account
(full black lines) and the situation in which the off-diagonal polarization
function ${\jnomatrix}_{\rho^+\pi^+}$ in Eq.~(\ref{rhopi}) is set to zero
(dashed green lines). The values correspond to the case in which the
four-fermion couplings are kept constant; similar results are found for
$B$-dependent couplings. It is seen that, as expected, the mixing leads to a
``repulsion'' between the $\tilde \pi^+$ and $\tilde \rho^+$ states: the
energy of $\tilde \pi^+$ is reduced, while that of the $\tilde \rho^+$
becomes enhanced. The repulsion gets larger as $eB$ increases, reaching an
effect of about $20\%$ for the $\tilde\pi^+$ energy at $eB =  1$~GeV$^2$.

\begin{figure}[h]
\centering{}\includegraphics[width=0.7\textwidth]{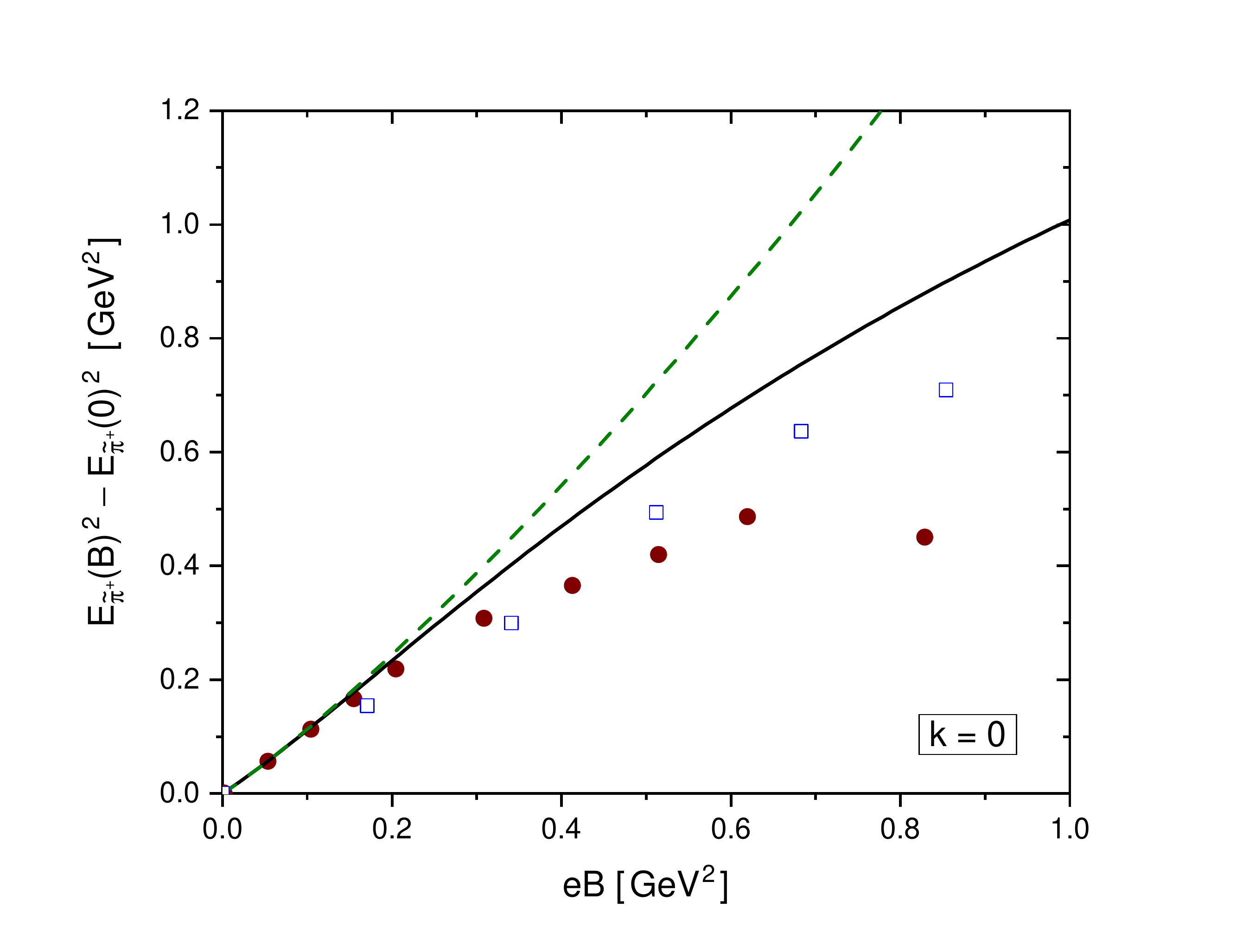}
\caption{(Color online) Squared energy of the $\tilde\pi^+$ mass eigenstate
for the Landau mode $k=0$ and vanishing component of the momentum in the
direction of $\vec B$. Values are given with respect to the squared mass for
vanishing external field. Solid and dashed lines correspond to the
calculations with and without the inclusion of the $\rho^+ - \pi^+$ mixing
terms, respectively. For comparison, lattice QCD data from
Ref.~\cite{Andreichikov:2016ayj} (open blue squares) and
Ref.~\cite{Ding:2020hxw} (full brown circles) are also shown.}
\label{Fig4}
\end{figure}

Finally, in Fig.~\ref{Fig4} we compare our results for $\tilde\pi^+$
energies with those obtained in lattice QCD analyses. The curves show the
values of squared $E_{\pi^+}$ energies with respect to $B=0$ squared masses,
considering both our numerical calculations with (full black line) and
without (dashed green line) pseudoscalar-vector meson mixing. As in
Fig.~\ref{Fig3}, the plots correspond to the case in which $g_s$ and $g_v$
do not depend on $B$. Open blue squares correspond to lattice QCD results
from Ref.~\cite{Andreichikov:2016ayj}, obtained using quenched Wilson
fermions and $m_\pi(B=0)= 395$~MeV, while full brown circles correspond to
the simulations reported in Ref.~\cite{Ding:2020hxw}, which were performed
using a highly improved staggered quark action with $m_\pi(B=0)= 220$~MeV.
We observe that the incorporation of the $\pi^+ - \rho^+$ mixing improves
the agreement between NJL model and LQCD results. However, it is seen that
the effect is not strong enough so as to account for the non-monotonous
behavior shown by the data from Ref.~\cite{Ding:2020hxw} for large values of
the magnetic field. Regarding the $\tilde\rho^+$ state, lattice results show
some variation depending on the lattice spacing and the simulation method
(see e.g.\
Refs.~\cite{Andreichikov:2016ayj,Luschevskaya:2016epp,Bali:2017ian}). In any
case, it is found that in general the $\tilde\rho^+$ energy shows an
increasing behavior with the magnetic field, in qualitative agreement with
our results in the right panel of Fig.~\ref{Fig2}. The results are
found to be similar for the case of $B$ dependent coupling constants.

\section{Summary and conclusions}

In this work we have studied the mass spectrum of $\pi^+$ and $\rho^+$
mesons in the presence of an external uniform magnetic field $\vec{B}$. This
has been done in the framework of a two-flavor NJL-like model that includes
scalar, pseudoscalar and vector four-fermion couplings. Due to the presence
of Schwinger phases, which induce the breakdown of translational invariance
in quark propagators, it is seen that charged meson polarization functions
do not become diagonal in the momentum basis. Here we have performed the
calculation of $\pi^+$ and $\rho^+$ polarization functions using an
expansion of the meson fields in terms of the solutions of the equations of
motion in the presence of the magnetic field. To account for the ultraviolet
divergences that usually arise in NJL-like models, we have considered a
magnetic field independent regularization (MFIR), which has been shown to
reduce the dependence of the results on the model parameters. Concerning the
effective coupling constants of the model, we have considered both the case
in which these parameters are fixed and the one in which they depend on the
external magnetic field.

In the case of the $\rho^+$ meson, our numerical calculations show that its
lowest energy state, which corresponds to a Landau level $k=-1$, lies above
$\sim 500$ MeV for values of $eB$ up to 1~GeV$^2$, both for the cases of
fixed and $B$-dependent couplings. In this way, our results ---which improve
upon previous two-flavor NJL model calculations that use a plane-wave
approximation for charged meson wavefunctions--- are not compatible with the
existence of a charged vector meson condensate induced by the magnetic
field. It is found that the $\rho^+$ state has a lower energy in the case of
$B$-dependent couplings, which leads to a better agreement with the results
from lattice QCD calculations.

Concerning the $\pi^+$ meson, it is seen that its lowest energy state, which
corresponds to the Landau level $k=0$, gets mixed with the corresponding
$\rho^+$ state for nonzero $B$. Our numerical results, both for the cases of
constant and $B$-dependent couplings, show that the mixing softens the
increase of the energy $E_{\tilde\pi^+}$ as a function of the magnetic
field, leading to energy values that lie slightly above those obtained for a
point-like particle. This softening effect is found to be favored by a
comparison with lattice QCD results.

For simplicity, in the present work we have not taken into account
axial-vector interactions. We expect to address their effect in a future
publication.

\section*{Acknowledments}

NNS would like to thank the Department of Theoretical Physics of the
University of Valencia, where part of this work has been carried out, for
their hospitality within the Visiting Professor program of the University of
Valencia. This work has been partially funded by CONICET (Argentina) under
Grants No.\ PIP17-700, by ANPCyT (Argentina) under Grant No.\ PICT17-03-0571,
PICT19-0792 and PICT20-01847, by the National University of La Plata (Argentina),
Project No.\ X824, by Ministerio de Ciencia e Innovaci\'on and Agencia
Estatal de Investigaci\'on (Spain), and European Regional Development Fund
Grant PID2019-105439 GB-C21, by EU Horizon 2020 Grant No.\ 824093
(STRONG-2020), and by Conselleria de Innovaci\'on, Universidades, Ciencia y
Sociedad Digital, Generalitat Valenciana GVA PROMETEO/2021/083.

\section*{Appendix A: Polarization functions at $B=0$}

\newcounter{eraAA}
\renewcommand{\thesection}{\Alph{eraAA}}
\renewcommand{\theequation}{\Alph{eraAA}\arabic{equation}}
\setcounter{eraAA}{1} \setcounter{equation}{0} 
\label{appa}

In this Appendix we provide the expressions of the regularized polarization
functions ${\jnomatrix}_{\pi}^{0,\rm reg}(q^2)$ and
${\jnomatrix}_{\rho}^{0,\rm reg}(q^2)$, obtained in the limit
$B=0$~\cite{Klimt:1989pm}. Notice that the mixing polarization functions
${\jnomatrix}_{\rho^+_\mu\pi^+} (\bar q)$ and
${\jnomatrix}_{\pi^+\rho^+_\mu} (\bar q)$ are zero in this limit. One has
\begin{eqnarray}
{\jnomatrix}_{\pi}^{0,\rm reg}(q^2) & = & -  2 N_c \Big[ I_{1}^{\rm reg} + q^2 I_{2}^{\rm reg}(q^2) \Big] \ \ ,
\nonumber \\
J_{\rho}^{0,\rm reg}(q^2) &=& \ \frac{4N_c}{3}\, \Big[(2M^2-q^2)I_{2}^{\rm
reg}(q^2)-2M^2 I_{2}^{\rm reg}(0)\Big]\ ,
\label{b0reg}
\end{eqnarray}
where $I_{1}^{\rm reg}$ and $I_{2}^{\rm reg}(q^2)$ are regularized
expressions of the integrals
\begin{eqnarray}
I_{1} & = & 4 \int \dfrac{d^4 p}{(2\pi)^4}\  \frac{1}{p^2+M^2}\ ,
\nonumber \\
I_{2}(q^2) & = & -2\int \dfrac{d^4 p}{(2\pi)^4}\ \frac{1}{[(p+q/2)^2+M^2]\,
[(p-q/2)^2+M^2]}\ .
\label{eqI1eI2}
\end{eqnarray}
Within the 3D-cutoff regularization scheme used in this work, the first of
these integrals is given by~\cite{Klimt:1989pm,Klevansky:1992qe}
\begin{equation}
I_{1}^{\rm reg} \ = \ \dfrac{1}{2 \pi^2} \left[ \Lambda^2\, r_\Lambda +
M^2 \ln\left( \dfrac{M}{\Lambda\,(1 + r_\Lambda)} \right) \right] \ ,
\label{I1freg}
\end{equation}
where we have defined $r_\Lambda = \sqrt{1 + M^2/\Lambda^2}$. In
the case of $I_{2}(q^2)$, we note that in order to determine the meson
masses, the external momentum $q$ has to be extended to the region $q^2 <
0$. Hence, we find it convenient to write $q^2=-m^2$, where $m$ is a
positive real number. Then, within the 3D-cutoff regularization scheme, the
regularized real part of $I_{2}(-m^2)$ can be written as~\cite{Klimt:1989pm}
\begin{align}
{\rm Re}\left[ I_{2}^{\rm reg}(-m^2)\right] = -\dfrac{1}{4\pi^2} &
 \left[ \arcsinh \left(\dfrac{\Lambda}{M}\right) - F(m^2)\right]  \ ,
\end{align}
where
\begin{eqnarray}
F(m^2) \ = \
\left\{
\begin{array}{ll}
  \sqrt{4 M^2/m^2 -1} \ \arctan \left( \dfrac{1}{r_\Lambda\,\sqrt{4 M^2/m^2 -1}} \right) & \mbox{\ \ if \ \ } m^2 < 4 M^2 \\
  \sqrt{1-4 M^2/m^2}  \ \arccoth \left( \dfrac{1}{r_\Lambda\,\sqrt{1-4 M^2/m^2}} \right) & \mbox{\ \ if \ \ }
  4 M^2 < m^2 < 4(M^2+\Lambda^2) \\
  \sqrt{1-4 M^2/m^2}  \ \arctanh \left( \dfrac{1}{r_\Lambda\,\sqrt{1-4 M^2/m^2}} \right) & \mbox{\ \ if \ \ }
  m^2 > 4(M^2+\Lambda^2)
\end{array}
\right. \ . \nonumber
\end{eqnarray}
For the regularized imaginary part we get
\begin{equation}
{\rm Im} \left[ I_{2}^{\rm reg}(-m^2)\right] \ = \
\left\{
\begin{array}{cl}
  -\dfrac{1}{8 \pi} \sqrt{1-4 M^2/m^2} & \mbox{\ \ if \ \ } 4 M^2 < m^2 < 4(M^2+\Lambda^2) \\
 \rule{0cm}{0.79cm} 0 & \mbox{\ \ otherwise}\\
\end{array}
\right. \ .
\label{ima2}
\end{equation}

\section*{Appendix B: Vector mesons in an external magnetic field.}

\newcounter{eraB}
\renewcommand{\thesection}{\Alph{eraB}}
\renewcommand{\theequation}{\Alph{eraB}\arabic{equation}}
\setcounter{eraB}{2} \setcounter{equation}{0} 
\label{appb}

In this Appendix we show that the functions introduced in
Eq.~(\ref{Ritusvec}) correspond to solutions of the equations of
motion of a charged vector meson in the presence of a constant magnetic
field, provided the associated dispersion relation
\begin{eqnarray}
E^2 \ = \ - q_4^2 \ = \ m^2 + (2 k +1) B_{\qindex} + q_3^2
\label{energy}
\end{eqnarray}
is satisfied.

We start from the equation of motion for a spin 1 field given in
Ref.~\cite{Corben:1940zz}. In Euclidean space one has
\begin{eqnarray}
\left[ \left( D_\alpha D_\alpha - m^2 \right) \delta_{\mu\nu} + 2 i Q F_{\mu\nu} \right] V_\nu(x)
\ = \ 0\ ,
\label{equ}
\end{eqnarray}
which has to be supplemented by the transversality condition
\begin{eqnarray}
D_\mu\, V_\mu(x) \ = \ 0 \ .
\label{trans}
\end{eqnarray}
In these equations, $Q$ stands for the electric charge of the vector field
$V_\mu(x)$, the covariant derivative $D_\alpha$ is given by $D_\alpha\ = \
\partial_{\alpha}-i\, Q \mathcal{A}_{\alpha}$, and $F_{\mu\nu} =
\partial_\mu \mathcal{A}_\nu - \partial_\nu \mathcal{A}_\mu$. For the
particular case of constant magnetic field along the $z$-axis, using the
Landau gauge one has $\mathcal{A}_\mu = B\, x_1\, \delta_{\mu 2}$, and
Eq.~(\ref{equ}) reduces to
\begin{eqnarray}
\left( \mathbbm{D}_{\mu\nu} - m^2 \delta_{\mu\nu} \right) V_\nu(x) \ = \ 0\ ,
\end{eqnarray}
where $\mathbbm{D}$ is a $4\times 4$ matrix given by
\begin{eqnarray}
\mathbbm{D} = \left[ \left( \nabla_1^2 + (\nabla_2 - i s B_{\qindex} x_1)^2 + \nabla_3^2 + \nabla_4^2 \right)
\left(%
\begin{array}{cc}
  \mathbbm{1} & 0\\
  0& \mathbbm{1} \\
\end{array}%
\right)
+ 2 s B_Q
\left(%
\begin{array}{cc}
 \sigma_2 & 0 \\
  0 & 0 \\
\end{array}%
\right)
\right]\ ,
\label{operator}
\end{eqnarray}
where $s= \mbox{sign}(Q B)$, $B_{\small Q}=| Q B|$ and $\sigma_2$ is a Pauli
matrix. Note that each entry in the matrices appearing in this equation
should be understood as a $2\times 2$ matrix, with $\mathbbm{1} =
\mbox{diag}(1,1)$.

Next, let us consider a function of the form introduced in
Eq.~(\ref{Ritusvec}), namely
\begin{eqnarray}
V_\mu(x) \ = \ \rhofunmat_{\mu\nu}(x,\bar q) \, e_{\nu}(\bar q)\ ,
\label{vq}
\end{eqnarray}
with $\bar q = (k,q_2,q_3,q_4)$. As in the main text, the functions
$\rhofunmat_{\mu\nu}(x,\bar q)$ are defined as
\begin{equation}
\rhofunmat_{\mu\nu}(x,\bar q) \ = \ \sum_{\ell = -1}^1 \rhofun_{\ell}(x,\bar q)\,
\Delta^{(\ell)}_{\mu\nu}\ ,
\label{rhofun}
\end{equation}
where
\begin{equation}
\rhofun_{\ell}(x,\bar q) \ = \ N_{k-s\ell} \; e^{i (q_2 x_2 + q_3 x_3 + q_4 x_4)}
\, D_{k-s\ell}(r)\ .
\label{Gq}
\end{equation}
Here, $D_{n}(x)$ are the cylindrical parabolic functions, with the
convention $D_n(x) =0$ if $n <0$, and we have used the definitions $N_n=
(4\pi B_{\qindex})^{1/4}/\sqrt{n!}$ and $r = s \sqrt{2/B_{\qindex}}(s
B_{\qindex}\, x_1-q_2)$. The $4\times 4$ matrices $\Delta^{(\ell)}$, $\ell
=-1,0,1$ are given by
\begin{eqnarray}
\Delta^{(1)} = \frac{1}{\sqrt 2}
\left(%
\begin{array}{cccc}
  0 & 1 & 0 & 0 \\
  0 & -i & 0 & 0 \\
  0 & 0 & 0 & 0 \\
  0 & 0 & 0 & 0 \\
\end{array}%
\right)
\, , \quad
\Delta^{(0)} =
\left(%
\begin{array}{cccc}
  0 & 0 & 0 & 0 \\
  0 & 0 & 0 & 0 \\
  0 & 0 & 1 & 0 \\
  0 & 0 & 0 & 1 \\
\end{array}%
\right)
\, , \quad
\Delta^{(-1)} = \frac{1}{\sqrt 2}
\left(%
\begin{array}{cccc}
  1 & 0 & 0 & 0 \\
  i & 0 & 0 & 0 \\
  0 & 0 & 0 & 0 \\
  0 & 0 & 0 & 0 \\
\end{array}%
\right)\, .
\end{eqnarray}
The election of these matrices is not unique; the above form has been chosen
taking into account that the operator in Eq.~(\ref{operator}) is diagonal in
the (3,4) subspace, while it leads to a mixing between components 1 and 2.
Note that, in order to have non-vanishing solutions, we must have $k-s \ell
\geq 0$. Given the possible values of $\ell$ ($=0,\pm 1$) and $s$ ($=\pm
1$), this implies that $k \geq -1$.

Using the explicit form of $\rhofunmat_{\mu\nu}(x,\bar q)$, it is
not difficult to prove the relation
\begin{eqnarray}
\mathbbm{D}_{\alpha\beta} \, \rhofunmat_{\beta\gamma}(x,\bar q) \
= \ - \left[ (2 k+1) B_{\qindex} + q_3^2 + q_4^2 \right]
\rhofunmat_{\alpha\gamma}(x,\bar q)\ .
\end{eqnarray}
In this way, it follows that the functions $V_\mu(x)$ in Eq.~(\ref{vq}) are
solutions of Eq.~(\ref{equ}), provided Eq.~(\ref{energy}) is satisfied. In
fact, these functions are equivalent to those introduced by
Ritus~\cite{Ritus:1978cj} for the case of spin $1/2$ fermions.

To determine the set of vectors $e_{\nu}(\bar q)$ that satisfy the
transversality condition in Eq.~(\ref{trans}), it is convenient to consider
the identity
\begin{eqnarray}
D_\mu \, \rhofunmat_{\mu\nu}(x,\bar q) \ = \ i \,
\rhofun_0(x,\bar q) \, [\Pi_\nu(\bar q)]^\ast\ ,
\label{relpinu}
\end{eqnarray}
where
\begin{eqnarray}
\Pi_\nu(\bar q) = \left(i s \sqrt{ B_{\qindex} k_-}\, , \, -i s \sqrt{  B_{\qindex} k_+}\,
, \, q_3 \, , \, q_4\right)\ ,
\label{pimu}
\end{eqnarray}
with $k_\pm = k + (1\mp s)/2$. From Eqs.~(\ref{vq}) and (\ref{relpinu}), the
transversality condition can be expressed as
\begin{eqnarray}
[\Pi_\nu(\bar q)]^\ast\, e_{\nu}(\bar q) \ = \ 0 \ .
\label{transvcond}
\end{eqnarray}
Note that $\Pi_\nu(\bar q)$ plays here the same role as the
four-momentum in the $B=0$ case. In fact, it is easy to see that
\begin{eqnarray}
\Pi^2\ \equiv \ [\Pi_\nu(\bar q)]^\ast \, \Pi_\nu(\bar q) \
= \ (2 k+1) B_{\qindex} + q_3^2 + q_4^2 \ ,
\label{pi2}
\end{eqnarray}
which implies that the condition in Eq.~(\ref{energy}) leads to $\Pi^2 =
-m^2$.

We denote by $\epsilon_{\nu}(\bar q,a)$ each of the independent
normalized solutions of Eq.~(\ref{transvcond}). They correspond to the
different possible polarization vectors of the spin 1 field. For $k \geq 1$
one can find three independent solutions. In that case, using the notation
$\bar q_{(k)} = (k,q_2,q_3,q_4)$, and taking for definiteness $s=+1$, we can
choose a basis formed by the vectors
\begin{eqnarray}
\epsilon_{\nu}(\bar q_{(k)},1) &=& \frac{\left(
   q_\parallel^2 \ , \ 0 \ , \
  i q_3 \sqrt{(k+1)B_{\qindex}} \ , \ i q_4  \sqrt{(k+1)B_{\qindex}}\right)}
{\sqrt{q_\parallel^2 \, [(k+1) B_{\qindex}+ q_\parallel^2]  }}\ ,
\nonumber \\[2.mm]
\epsilon_{\nu}(\bar q_{(k)},2) &=& \frac{
\left( 0 \ ,\ 0 \ ,\ iq_4\ ,\ -iq_3 \right)}{\sqrt{- q_\parallel^2}} \ ,
\nonumber  \\[2.mm]
\epsilon_{\nu}(\bar q_{(k)} ,3) &=& \frac{
\left(
  - B_{\qindex} \sqrt{k(k+1)}\ ,\
-[ (k+1) B_{\qindex} +q_\parallel^2]\ ,\
  i q_3 \sqrt{k B_{\qindex}} \ ,\
  i q_4 \sqrt{k B_{\qindex}} \right)}
{\sqrt{\Pi^2 \, [(k+1)B_{\qindex} + q_\parallel^2]}}\ ,
\label{polvect}
\end{eqnarray}
where $q_\parallel^2 = q_3^2 + q_4^2$. For $s=-1$, the corresponding results
can be obtained by exchange of the first two components of these vectors.

Due to the restrictions imposed by the condition $k-s \ell \geq 0$, the
situations for $k=-1$ and $k=0$ have to be considered separately. In the
case $k=-1$, from Eqs.~(\ref{rhofun}) and (\ref{Gq}) it is seen that
only one independent solution of the form given by Eq.~(\ref{vq}) can be
constructed. The associated polarization vector is
\begin{eqnarray}
\epsilon_\nu(\bar q_{(-1)} ,1) \ = \ \left( 1 \ , \ 0 \ ,
\ 0 \ , \ 0 \right)
\label{polm1}
\end{eqnarray}
for $s=1$, and $\epsilon_\nu(\bar q_{(-1)} ,1) = (0,1,0,0)$ for $s=-1$. On
the other hand, for $k=0$ two independent transverse solutions can be
constructed. In this case, a suitable choice for the polarization vectors
is
\begin{eqnarray}
\epsilon_\nu(\bar q_{(0)},1) &=&
\left(
  \delta_{1s} \ q_\parallel^2 \  , \
 \delta_{-1s} \ q_\parallel^2 \  , \
  i q_3 \sqrt{B_{\qindex}} , \
 i q_4  \sqrt{B_{\qindex}}
\right)/ \sqrt{q_\parallel^2 (q_\parallel^2 + B_{\qindex}) } \ ,
\nonumber \\[2.mm]
\epsilon_\nu(\bar q_{(0)},2) &=&
\left(  0 \ , \ 0 \ , \ iq_4 \ , \ -iq_3\right)/{\sqrt{- q_\parallel^2}} \ .
\label{polk0}
\end{eqnarray}

Replacing the polarization vectors (\ref{polvect}) in Eq.~(\ref{vq}), and
using the on-shell condition Eq.~(\ref{energy}), one recovers the known
solutions for a vector boson in a constant magnetic field (see e.g.
Ref.~\cite{Nikishov:2001fd}) written in Euclidean space.

Finally, note that for $k\geq 0$ an extra ``longitudinal'' polarization
vector $\epsilon_\nu(\bar q_{(k)} ,4)$ can be defined as
\begin{equation}
\epsilon_\nu(\bar q_{(k)},4) \ = \ \Pi_\nu(\bar q)\,
/\,{\sqrt{- \Pi^2}} \ .
\end{equation}
In the case $k=-1$ the relation in Eq.~(\ref{trans}) is always satisfied.
Therefore, no ``longitudinal'' polarization can be constructed.

\section*{Appendix C: Analytic continuation of polarization functions}

\newcounter{eraC}
\renewcommand{\thesection}{\Alph{eraC}}
\renewcommand{\theequation}{\Alph{eraC}\arabic{equation}}
\setcounter{eraC}{3} \setcounter{equation}{0} 

In this Appendix we discuss how to evaluate the magnetic contributions to
the polarization functions for energies beyond the threshold of $2M$. The
integrals to be analyzed are those given by Eqs.~(\ref{rhorhom1}),
(\ref{rhopi}), (\ref{Jmagpipi}) and (\ref{Jmagrhorho}).

Let us start by considering the integral in Eq.~(\ref{rhorhom1}), which
corresponds to the $\rho^+$ polarization function for the $k=-1$ Landau
mode. It is convenient to separate this integral into ultraviolet and
infrared pieces, namely
\begin{equation}
{\jnomatrix}_{\rho^+\!\rho^+}^{\rm mag}(-1,-m^2) \ = \ J_{\rho^+\!\rho^+}^{\rm uv}(-1,-m^2)
\, + \, J_{\rho^+\!\rho^+}^{(B)\,{\rm ir}}(-1,-m^2)\, + \, J_{\rho^+\!\rho^+}^{(0)\,{\rm ir}}(-m^2)\ ,
\end{equation}
where
\begin{eqnarray}
{\jnomatrix}_{\rho^+\!\rho^+}^{\rm uv}(-1,-m^2) & = &
-\frac{N_c}{4\pi^2}\,\int_{-1}^1 dv \int_0^{4/m^2} dz
\ e^{-z [M^2 - (1-v^2)m^2/4]} \nonumber \\
& & \times \, \bigg\{ \frac{(1+t_u)\,(1+t_d)}{\alpha_+}
\,\Big[ M^2 + \frac{1}{z} + \frac{1-v^2}{4}\,(m^2 - B_e)\Big]
e^{-z (1-v^2)B_e/4} \nonumber \\
& & \hspace{0.5cm} - \, \frac{1}{z}
\,\Big[ M^2 + \frac{1}{z} + \frac{1-v^2}{4}\,m^2\Big] \bigg\}\ , \nonumber \\
{\jnomatrix}_{\rho^+\!\rho^+}^{(B)\,{\rm ir}}(-1,-m^2) & = &
-\frac{N_c}{4\pi^2}\, \int_{-1}^1 dv \int_{4/m^2}^\infty dz
\ e^{-z [M^2 - (1-v^2)(m^2-B_e)/4]} \nonumber \\
& & \times \, \frac{(1+t_u)\,(1+t_d)}{\alpha_+}
\,\Big[ M^2 + \frac{1}{z} + \frac{1-v^2}{4}\,(m^2 - B_e)\Big]\ , \nonumber \\
{\jnomatrix}_{\rho^+\!\rho^+}^{(0)\,{\rm ir}}(-m^2) & = &
\frac{N_c}{4\pi^2}\, \int_{-1}^1 dv \int_{4/m^2}^\infty \frac{dz}{z}
\ e^{-z [M^2 - (1-v^2)m^2/4]}\,\Big( M^2 + \frac{1}{z} + \frac{1-v^2}{4}\,m^2\Big)\ .\ \
\end{eqnarray}
It is easy to see that the threshold for the appearance of absorptive parts
for $J_{\rho^+\!\rho^+}^{\rm uv}(-1,-m^2)$ and
$J_{\rho^+\!\rho^+}^{(B)\,{\rm ir}}(-1,-m^2)$ is given by $m_{th}^{(-1)}=
\sqrt{4 M^2 + B_e}$. On the other hand, the $J_{\rho^+\!\rho^+}^{(0)\,{\rm
ir}}(-1,-m^2)$ is divergent for $m^2\geq 4M^2$. To go beyond this limit, one
can perform an analytic continuation. It can be seen that after integration
over $z$ one gets
\begin{equation}
{\jnomatrix}_{\rho^+\!\rho^+}^{(0)\,{\rm ir}}(-m^2) \ = \
\frac{N_c\, m^2}{8\pi^2}\bigg[\frac{\sqrt{\pi}}{2}\; {\rm
erf}(1)\, \exp\big(\beta^2\big)\, + \,
\int_{-1}^1 dv \,(1-v^2)\,E_1(v^2-\beta^2)\,\bigg]\ ,
\end{equation}
where $\beta^2 = 1-4M^2/m^2$, ${\rm erf}(x)$ is the error function, and
$E_1(x)$ is the exponential integral, which can be written as
\begin{equation}
E_1(x) \ = \ - \gamma - \ln x + E_{in}(x)\ ,
\label{e1x}
\end{equation}
with $E_{in}(x) =\sum_{k=1}^\infty (-1)^{k+1} x^k/ (k!\, k)$. For $m^2$
larger than $4M^2$ (i.e., $\beta^2 >0$), the logarithm in Eq.~(\ref{e1x})
can be extended as $\ln (x-i\epsilon) = \ln |x|-i\,\pi$ for negative values of
$x$. This leads to a finite expression for
${\jnomatrix}_{\rho^+\!\rho^+}^{(0)\,{\rm ir}}(-m^2)$ that includes an
imaginary part
\begin{equation}
{\rm Im}\Big[{\jnomatrix}_{\rho^+\!\rho^+}^{(0)\,{\rm ir}}(-m^2)\Big] \ = \
\frac{N_c\, m^2}{8\pi}\, \int_{-\beta}^\beta dv \,(1-v^2) \ = \
\frac{N_c}{6\pi}\,\beta\,(m^2+2M^2)\ ,
\end{equation}
which cancels exactly with the imaginary part arising from the regularized
$B=0$ piece of the polarization function $\jnomatrix^{0,\rm
reg}_{\rho}(-m^2)$, see Eqs.~(\ref{b0reg}) and (\ref{ima2}). Thus, it is
seen that the threshold $m^2= 4M^2$ is only apparent, the actual threshold
for quark-antiquark pair production in this case being located at
$m_{th}^{(-1)}$.

The situation is similar in the case of the $k=0$ Landau mode. However, the
corresponding quark-antiquark production threshold $m_{th}^{(0)}$ is lower
than $m_{th}^{(-1)}$; hence, it is interesting to obtain the expressions for
the analytic continuation of the polarization functions even beyond this
limit. Let us consider the function ${\jnomatrix}_{\rho^+\!\rho^+}^{\rm mag}
(0,-m^2)$, given by Eq.~(\ref{Jmagrhorho}). It is convenient to separate it
into four terms, namely
\begin{equation}
{\jnomatrix}_{\rho^+\!\rho^+}^{\rm mag}(0,-m^2) \ = \ J_{\rho^+\!\rho^+}^{\rm uv}(0,-m^2)
\, + \, J_{\rho^+\!\rho^+}^{(B1)\,{\rm ir}}(0,-m^2)\, + \, J_{\rho^+\!\rho^+}^{(B2)\,{\rm ir}}(0,-m^2)\,
+ \, J_{\rho^+\!\rho^+}^{(0)\,{\rm ir}}(-m^2)\ ,
\end{equation}
where
\begin{eqnarray}
{\jnomatrix}_{\rho^+\!\rho^+}^{\rm uv}(0,-m^2) & = & -\dfrac{N_c}{4\pi^2}
 \int_{-1}^1 dv\ \bigg\{ \int_0^{4/(m^2+B_e)} dz \ e^{-z[M^2 - (1-v^2)(m^2+B_e)/4]}
 \nonumber \\
 & & \times\,\left[\, \dfrac{(1-t_u \,t_d)}{\alpha_+} \bigg( M^2 + \frac{1-v^2}{4}\, (m^2+B_e) \bigg)
 + \dfrac{(1-t_u^2)\,(1-t_d^2)}{\alpha_+^{2}}\,\right]
\nonumber \\
& &
- \,\int_0^{4/m^2} \dfrac{dz}{z} \ e^{-z[M^2 - (1-v^2)m^2/4]}
\bigg( M^2 + \dfrac{1}{z} + \frac{1-v^2}{4}\, m^2 \bigg) \bigg\}\ ,
\nonumber \\
{\jnomatrix}_{\rho^+\!\rho^+}^{(B1)\,{\rm ir}} (0,-m^2) & = & -\dfrac{N_c}{4\pi^2}
 \int_{-1}^1 dv\ \int_{4/(m^2+B_e)}^\infty dz \ e^{-z[M^2 - (1-v^2)(m^2+B_e)/4]}
 \nonumber \\
 & & \times\,\bigg[ \left(\, \dfrac{(1-t_u \,t_d)}{\alpha_+} - \frac{2B_e}{9}\,
 e^{-z(1+v)B_e/3} \right)
   \bigg( M^2 + \frac{1-v^2}{4}\, (m^2+B_e) \bigg) \nonumber \\
 & & +\; \dfrac{(1-t_u^2)\,(1-t_d^2)}{\alpha_+^{2}}\,\bigg] \ , \nonumber \\
{\jnomatrix}_{\rho^+\!\rho^+}^{(B2)\,{\rm ir}} (0,-m^2) & = & -\,\dfrac{N_c\,B_e}{18\pi^2}
 \int_{-1}^1 dv\ \int_{4/(m^2+B_e)}^\infty dz \ e^{-z[M^2 - (1-v^2)(m^2+B_e)/4 + (1+v)B_e/3]}
 \nonumber \\
 & & \times \, \bigg[ M^2 + \frac{1-v^2}{4}\, (m^2+B_e) \bigg]\ ,
\end{eqnarray}
while $J_{\rho^+\!\rho^+}^{(0)\,{\rm ir}}(-m^2)$ is the same function
analyzed in the $k=-1$ case (and the cancellation of its imaginary part
proceeds in the same way as discussed above). After some analysis, it can be
shown that the integrals in $J_{\rho^+\!\rho^+}^{(B2)\,{\rm ir}}(0,-m^2)$
are convergent for $m^2 < m_{th}^{(0)\;2} = (M + \sqrt{M^2+2B_e/3})^2-B_e$,
whereas for $J_{\rho^+\!\rho^+}^{(B1)\,{\rm ir}}(0,-m^2)$ the region of
convergence extends up to $m_{th}^{(0)\,\prime\;2} = (M +
\sqrt{M^2+4B_e/3})^2-B_e$. In what follows we discuss how to perform an
analytic extension of $J_{\rho^+\!\rho^+}^{(B2)\,{\rm ir}}(0,-m^2)$ in order
to get a definite result for the polarization function between these two
thresholds. After integration over $z$ one gets
\begin{equation}
{\jnomatrix}_{\rho^+\!\rho^+}^{(B2)\,{\rm ir}} (0,-m^2) \ = \ -\,\dfrac{N_c\,B_e}{18\pi^2}
 \int_{-1}^1 dv\ \frac{4\,r_0^2 + 1-v^2}{(v+\delta)^2+\lambda^2}\; e^{-(v+\delta)^2-\lambda^2}\ ,
\label{jb2ir}
\end{equation}
where we have introduced the definitions
\begin{equation}
r_0 = \frac{M}{\sqrt{m^2+B_e}}\ , \qquad r_d =
\sqrt{\frac{M^2+2B_e/3}{m^2+B_e}}\ , \qquad \delta = r_d^2 - r_0^2
\end{equation}
and
\begin{equation}
\lambda^2 \ = \ \Big[(r_d+r_0)^2-1\Big]\,\Big[1-(r_d-r_0)^2\Big]\ .
\end{equation}
The expression in Eq.~(\ref{jb2ir}) can be written as
\begin{eqnarray}
{\jnomatrix}_{\rho^+\!\rho^+}^{(B2)\,{\rm ir}} (0,-m^2) & = & \dfrac{N_c\,B_e}{18\pi^2}
\,\Bigg\{\frac{\sqrt{\pi}}{2}\,e^{-\lambda^2}\,\Big[{\rm erf}(1+\delta)+{\rm erf}(1-\delta)\Big]\,
 + \, \delta\,\Big[E_1(4\,r_d^2)-E_1(4\,r_0^2)\Big]\nonumber \\
& & +\,2(1-\delta+\lambda^2)\left[
\int_{-1}^1 dv\ \frac{1-e^{-(v+\delta)^2-\lambda^2}}{(v+\delta)^2+\lambda^2}
\,+\,F(m^2+B_e)\right]\Bigg\}\ ,
\end{eqnarray}
where
\begin{equation}
F(m^2+B_e)\ = \
\int_{-1}^1 dv\ \frac{1}{(v+\delta)^2+\lambda^2}\ .
\label{fmbe}
\end{equation}
For $m < m_{th}^{(0)}$ one has $\lambda^2>0$ and the integral in Eq.~(\ref{fmbe}) can be
done explicitly, leading to
\begin{equation}
F(m^2+B_e)_{m\, < \, m_{th}^{(0)}}\ = \ \frac{1}{\lambda}\left[
\arctan\bigg(\frac{\lambda}{1+\delta}\bigg) \,
+ \, \arctan\bigg(\frac{\lambda}{1+\delta}\bigg) \, -\, \pi\right]\ .
\label{fmbe2}
\end{equation}
On the other hand, for $m$ beyond the threshold $m_{th}^{(0)}$ one has
$\lambda^2<0$. Defining $\bar\lambda^2=-\lambda^2$, the function above can
be analytically extended to
\begin{equation}
F(m^2+B_e)_{m\, > \, m_{th}^{(0)}}\ = \ \frac{1}{\bar\lambda}\left[
\arctanh\bigg(\frac{\bar\lambda}{1+\delta}\bigg) \,
+ \, \arctanh\bigg(\frac{\bar\lambda}{1+\delta}\bigg) \, -\,i\, \pi\right]\ ,
\end{equation}
implying the existence of an absorptive part in the polarization function.
At the threshold one has $r_0+r_d=1$, thus $\lambda^2=0$ and
$J_{\rho^+\!\rho^+}^{(B2)\,{\rm ir}}(0,-m^2)$ is divergent.

A similar procedure can be carried out in the case of the magnetic piece of
the polarization function ${\jnomatrix}^{\rm reg}_{\pi^+\!\pi^+}(0,-m^2)$,
for $m<m_{th}^{(0)\,\prime}$. The corresponding expressions are found to be
given by
\begin{equation}
{\jnomatrix}_{\pi^+\!\pi^+}^{\rm mag}(0,-m^2) \ = \ J_{\pi^+\!\pi^+}^{\rm uv}(0,-m^2)
\, + \, J_{\pi^+\!\pi^+}^{(B1)\,{\rm ir}}(0,-m^2)\, + \, J_{\pi^+\!\pi^+}^{(B2)\,{\rm ir}}(0,-m^2)\,
+ \, J_{\pi^+\!\pi^+}^{(0)\,{\rm ir}}(-m^2)\ ,
\end{equation}
where
\begin{eqnarray}
{\jnomatrix}_{\pi^+\!\pi^+}^{\rm uv}(0,-m^2) & = & -\dfrac{N_c}{4\pi^2}
 \int_{-1}^1 dv\ \bigg\{ \int_0^{4/(m^2+B_e)} dz \ e^{-z[M^2 - (1-v^2)(m^2+B_e)/4]}
 \nonumber \\
 & & \times\,\left[\, \dfrac{(1-t_u \,t_d)}{\alpha_+} \bigg( M^2 + \,
 \frac{1}{z}\, + \frac{1-v^2}{4}\, (m^2+B_e) \bigg)
 + \dfrac{(1-t_u^2)\,(1-t_d^2)}{\alpha_+^{2}}\,\right]
\nonumber \\
& &
- \,\int_0^{4/m^2} \dfrac{dz}{z} \ e^{-z[M^2 - (1-v^2)m^2/4]}
\bigg( M^2 + \dfrac{2}{z} + \frac{1-v^2}{4}\, m^2 \bigg) \bigg\}\ ,
\nonumber \\
{\jnomatrix}_{\pi^+\!\pi^+}^{(B1)\,{\rm ir}} (0,-m^2) & = & -\dfrac{N_c}{4\pi^2}
 \int_{-1}^1 dv\ \int_{4/(m^2+B_e)}^\infty dz \ e^{-z[M^2 - (1-v^2)(m^2+B_e)/4]}
 \nonumber \\
 & & \times\,\bigg[ \left(\, \dfrac{(1-t_u \,t_d)}{\alpha_+} - \frac{2B_e}{9}\,
 e^{-z(1+v)B_e/3} \right)
   \bigg( M^2 + \frac{1}{z} + \frac{1-v^2}{4}\, (m^2+B_e) \bigg) \nonumber \\
 & & +\; \dfrac{(1-t_u^2)\,(1-t_d^2)}{\alpha_+^{2}}\,\bigg] \ , \nonumber \\
{\jnomatrix}_{\pi^+\!\pi^+}^{(B2)\,{\rm ir}} (0,-m^2) & = & \dfrac{N_c\,B_e}{18\pi^2}
\,\Bigg\{2\gamma - 4 + \frac{\sqrt{\pi}}{2}\,e^{-\lambda^2}\,\Big[{\rm erf}(1+\delta)+{\rm erf}(1-\delta)\Big]\,
+ 2\ln\big(4\,r_0\,r_d\big) \nonumber \\
& & +\, \int_{-1+\delta}^{1+\delta} dv\
\bigg[2(1-\delta+\lambda^2)\frac{1-e^{-(v^2+\lambda^2)}}{v^2+\lambda^2}\,
-\,E_{in}(v^2+\lambda^2)\bigg]
\nonumber \\
& & + \; \delta\,\Big[E_{in}(4\,r_d^2)-E_{in}(4\,r_0^2)\Big]\,
+\,2(1-\delta)\,F(m^2+B_e)\,\Bigg\}\ ,
\nonumber \\
{\jnomatrix}_{\pi^+\!\pi^+}^{(0)\,{\rm ir}}(-m^2) & = &
\frac{N_c\, m^2}{16\pi^2}\bigg\{\int_{-1}^1 dv \,(2+\beta^2-3v^2)\,
\big[-\gamma - \ln |v^2-\beta^2| + E_{in}(v^2-\beta^2)\Big]
\nonumber\\
& & + \; 2\sqrt{\pi}\; {\rm erf}(1)\, e^{\,\beta^2}
+\,i\,4\pi\beta\,\theta(\beta^2)\bigg\} \ .
\end{eqnarray}
As in the case of the polarization function
${\jnomatrix}_{\rho^+\!\rho^+}^{\rm reg} (0,-m^2)$, for $m>2M$ the imaginary
part in ${\jnomatrix}_{\pi^+\!\pi^+}^{(0)\,{\rm ir}}(-m^2)$ cancels with the
imaginary part arising from $\jnomatrix^{0,\rm reg}_{\pi}(-m^2)$, whereas
for $m_{th}^{(0)} < m < m_{th}^{(0)\,\prime}$ one gets an absorptive part
coming from the function $F(m^2+B_e)$ in
${\jnomatrix}_{\pi^+\!\pi^+}^{(B2)\,{\rm ir}} (0,-m^2)$ (beyond
$m_{th}^{(0)\,\prime}$, another absorptive contribution will arise from
${\jnomatrix}_{\pi^+\!\pi^+}^{(B1)\,{\rm ir}} (0,-m^2)$). Finally, for the
mixing polarization function ${\jnomatrix}_{\rho^+\!\pi^+}(0,-m^2)$ (which
does not need regularization in the ultraviolet limit) we obtain
\begin{equation}
{\jnomatrix}_{\rho^+\!\pi^+}^{\rm mag}(0,-m^2) \ = \ J_{\rho^+\!\pi^+}^{\rm uv}(0,-m^2)
\, + \, J_{\rho^+\!\pi^+}^{(B1)\,{\rm ir}}(0,-m^2)\, +
\, J_{\rho^+\!\pi^+}^{(B2)\,{\rm ir}}(0,-m^2)\ ,
\end{equation}
where
\begin{eqnarray}
{\jnomatrix}_{\rho^+\!\pi^+}^{\rm uv}(0,-m^2) & = & \dfrac{N_c\,M\sqrt{m^2+B_e}}{4\pi^2}
 \int_{-1}^1 dv \int_0^{4/(m^2+B_e)} dz \ \dfrac{(t_u - t_d)}{\alpha_+}\,
 e^{-z[M^2 - (1-v^2)(m^2+B_e)/4]}\ ,
 \nonumber \\
{\jnomatrix}_{\rho^+\!\pi^+}^{(B1)\,{\rm ir}} (0,-m^2) & = & \dfrac{N_c\,M\sqrt{m^2+B_e}}{4\pi^2}
 \int_{-1}^1 dv\int_{4/(m^2+B_e)}^\infty dz \ e^{-z[M^2 - (1-v^2)(m^2+B_e)/4]}
 \nonumber \\
 & & \times\,\bigg[ \dfrac{t_u - t_d}{\alpha_+} - \frac{2B_e}{9}\,
 e^{-z(1+v)B_e/3} \bigg] \ ,\nonumber \\
{\jnomatrix}_{\rho^+\!\pi^+}^{(B2)\,{\rm ir}} (0,-m^2) & = &
-\,\dfrac{2N_c\,M B_e}{9\pi^2\sqrt{m^2+B_e}}
\left[
\int_{-1+\delta}^{1+\delta} dv\ \frac{1-e^{-(v^2+\lambda^2)}}{v^2+\lambda^2}
\,+\,F(m^2+B_e)\right]\ ,
\end{eqnarray}
with $F(m^2+B_e)$ given by Eqs.~(\ref{fmbe}) and (\ref{fmbe2}).

\end{document}